\documentclass[lettersize,journal]{IEEEtran}

\IEEEoverridecommandlockouts
\usepackage{amsmath,amsfonts}
\usepackage[ruled,linesnumbered]{algorithm2e}
\usepackage{amssymb}
\usepackage{algorithmic}
\usepackage{bibentry}
\usepackage{amsfonts}
\usepackage{subfigure}
\usepackage{multirow}
\usepackage{amsmath}
\usepackage{graphicx}
\usepackage{epsfig}
\usepackage{color}
\usepackage{epstopdf}
\usepackage{ragged2e}
\usepackage{rotating}
\usepackage{caption}
\usepackage{float}
\usepackage{multicol}
\usepackage{amssymb}
\usepackage{graphicx}
\usepackage{bbding}
\usepackage{balance}
\usepackage{microtype}
\usepackage{threeparttable}
\usepackage{subfigure}
\usepackage{booktabs} 
\usepackage{multirow}
\usepackage{amsmath}
\usepackage[many]{tcolorbox}
\usepackage{mathtools}
\usepackage{etoolbox}
\usepackage{cases}
\usepackage{enumitem}
\usepackage{xcolor}
\usepackage{subfigure}

\usepackage{mathtools}

\usepackage{amssymb}
\usepackage{amsthm}
\usepackage{color}
\usepackage{multirow}

\newcommand{\ourname}{{TokenRec}}
\newcommand{\ourvqname}{{MQ-Tokenizer}}

\newcommand{\cut}[1]{{}}

\newcommand{\vE}{{\mathbf{E}}}

\makeatletter
\let\@@span\span
\def\sp@n{\@@span\omit\advance\@multicnt\m@ne}
\makeatother

\newcommand{\bc}{\begin{center}}
\newcommand{\ec}{\end{center}}

\newcommand{\bdm}{\begin{displaymath}}
\newcommand{\edm}{\end{displaymath}}

\newcommand{\beq}{\begin{equation}}
\newcommand{\eeq}{\end{equation}}

\newcommand{\bfl}{\begin{flushleft}}
\newcommand{\efl}{\end{flushleft}}

\newcommand{\bt}{\begin{tabbing}}
\newcommand{\et}{\end{tabbing}}

\newcommand{\beqn}{\begin{align}}
\newcommand{\eeqn}{\end{align}}

\newcommand{\beqs}{\begin{align*}} 
\newcommand{\eeqs}{\end{align*}}  

\DeclareUnicodeCharacter{0301}{*************************************}

\begin{document}

\title{TokenRec: Learning to Tokenize ID for LLM-based Generative Recommendations}

\author{Haohao Qu, Wenqi Fan$^*$, Zihuai Zhao, Qing Li,~\IEEEmembership{Fellow,~IEEE}\\
\IEEEcompsocitemizethanks{
\IEEEcompsocthanksitem H. Qu, Z. Zhao, and Q. Li are with the Department of Computing, The Hong Kong Polytechnic University. E-mail: haohao.qu@connect.polyu.hk, Scofield.zzh@gmail.com, qing-prof.li@polyu.edu.hk.
\IEEEcompsocthanksitem W. Fan is with the Department of Computing (COMP) and Department of Management and Marketing (MM), The Hong Kong Polytechnic University. E-mail: wenqifan03@gmail.com
}
\thanks{*Corresponding author: Wenqi Fan}
}

\markboth{IEEE TRANSACTIONS ON KNOWLEDGE AND DATA ENGINEERING, SUBMISSION 2024}%
{Qu \MakeLowercase{\textit{et al.}}: TokenRec: Learning to Tokenize ID for LLM-based Generative Recommendations.}




\maketitle
\IEEEpeerreviewmaketitle

\begin{abstract}
\justifying
There is a growing interest in utilizing large language models (LLMs) to advance next-generation Recommender Systems (RecSys), driven by their outstanding language understanding and reasoning capabilities. In this scenario, tokenizing users and items becomes essential for ensuring seamless alignment of LLMs with recommendations. While studies have made progress in representing users and items using textual contents or latent representations, challenges remain in capturing high-order collaborative knowledge into discrete tokens compatible with LLMs and generalizing to unseen users/items. To address these challenges, we propose a novel framework called \textbf{TokenRec}, which introduces an effective ID tokenization strategy and an efficient retrieval paradigm for LLM-based recommendations. Our tokenization strategy involves quantizing the masked user/item representations learned from collaborative filtering into discrete tokens, thus achieving smooth incorporation of high-order collaborative knowledge and generalizable tokenization of users and items for LLM-based RecSys. Meanwhile, our generative retrieval paradigm is designed to efficiently recommend top-K items for users, eliminating the need for the time-consuming auto-regressive decoding and beam search processes used by LLMs, thus significantly reducing inference time. Comprehensive experiments validate the effectiveness of the proposed methods, demonstrating that TokenRec outperforms competitive benchmarks, including both traditional recommender systems and emerging LLM-based recommender systems. Codes and data are available at https://github.com/Quhaoh233/TokenRec.

\end{abstract}


\section{Introduction}
\IEEEPARstart{A}s a prominent branch in the data mining field, recommender systems (\textbf{RecSys}) serve as an indispensable and effective technique in addressing information overload problems and enriching user experience across diverse applications~\cite{ning2024cheatagent,chen2023fairly,zhao2024recommender}, such as e-commerce, job search, and social media platforms.
To provide personalized recommendations that accord with user preferences, one of the most representative techniques is collaborative filtering (CF), which aims to capture collaborative knowledge by modeling the history of user-item interactions~\cite{fan2019graph}. 
For example, as the most classic CF method, Matrix Factorization (MF)~\cite{koren2009matrix} decomposes the use-item matrix into two low-rank matrices (i.e., user\&item representations) to represent each user and item with a unique ID (namely ID indexing/tokenizing), and calculates matching scores to predict user behaviors and generate recommendations via the inner product between their representations.
Due to the superior ability in representation learning on graph-structured data, Graph Neural Networks (GNNs)  (e.g., LightGCN~\cite{he2020lightgcn} and GTN~\cite{fan2022graph}) have been recently employed to significantly enhance CF by capturing higher-order collaborative knowledge on user-item interaction graphs for recommendations in IDs manners.
The main idea of the existing methods is to obtain learnable representations (i.e., token embeddings) for discrete user\&item IDs (i.e., tokens). 

\begin{figure*}[htbp]
\centering
{\includegraphics[width=0.99\linewidth]{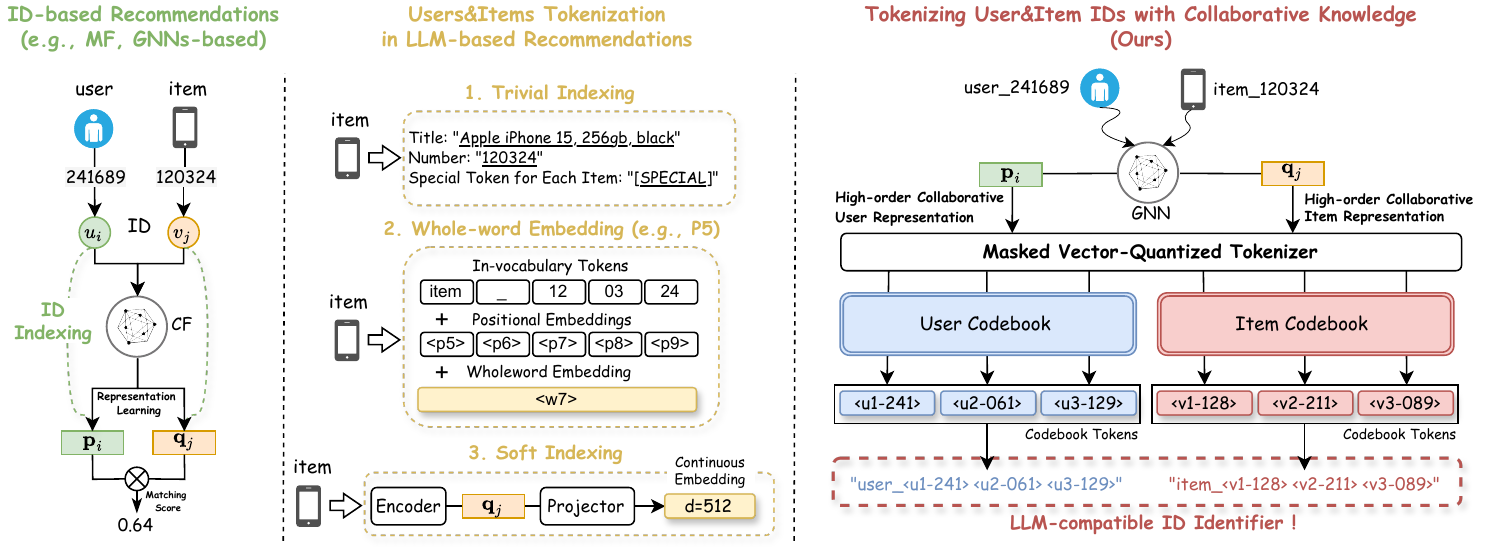}}
\vskip -0.1in
\caption{Comparison of ID tokenization methods in LLM-based recommendations. Unlike the existing methods, our approach can tokenize users and items with LLM-compatible tokens by leveraging high-order collaborative knowledge.
}
\vskip -0.1in
\label{fig:intro}
\end{figure*}

More recently, the rapid development of Large Language Models (LLMs) techniques (e.g., ChatGPT and LLaMA) has showcased notable milestones for revolutionizing natural language processing techniques~\cite{brown2020language,zhao2023survey, ning2025survey}.
Technically, the unprecedented power of LLMs can be attributed to the scaling up of model/parameter size alongside a tremendous amount of training corpus.
In particular, LLMs equipped with billion-scale parameters have exhibited unprecedented language understanding and generation abilities, along with remarkable generalization capability and in-context learning skills that facilitate LLMs to better generalize to unseen tasks and domains~\cite{zhao2023survey}. 
Given the emerging trends and aforementioned advancements of LLMs, LLM-empowered recommender systems have drawn increasing attention from recent studies and demonstrated distinctive abilities for advancing recommender systems~\cite{zhao2024recommender}. 
For example, ChatRec~\cite{gao2023chat} integrates ChatGPT into conversational recommendations, which enables users to deliver their explicit preferences in natural language. 
P5~\cite{geng2022recommendation} introduces an LLM-based recommendation framework that unifies diverse recommendation tasks by multi-task prompt-based pre-training.

Unlike the ID-based recommendation methods (e.g., MF and GNN-based), users\&items tokenization is one of the most critical steps to take advantage of LLMs within recommendations, as shown in Figure~\ref{fig:intro}. 
To be more specific, the naive approach known as Independent Indexing (IID) assigns special tokens (i.e., ID) to tokenize each user and item within language models directly. However, this approach becomes infeasible and unrealistic when dealing with large-scale real-world recommender systems, in which the sizes of users and items typically reach billions, dramatically expanding the token vocabulary in LLMs.
Meanwhile, as a natural solution, textual title indexing is proposed to utilize LLMs' in-vocabulary tokens to represent items based on their titles and descriptions, such as the example "\emph{Apple iPhone 15, 256 GB, black}", thus avoiding vocabulary size explosion in LLMs~\cite{gao2023chat,bao2023tallrec}. 
To achieve a closer alignment between recommendations and natural language, P5 employs \emph{whole-word embedding}~\cite{takase2020all} to indicate whether consecutive sub-word tokens originate from the same entities (i.e., user/item).
Some studies use continuous embedding (i.e., soft indexing) learned from encoders to represent users\&items in LLM-based recommendations~\cite{liao2023llara,zhang2023collm}.
Despite the success mentioned above, the majority of existing methods for users\&items tokenization for LLM-based recommender systems still have several limitations. 
For example, the use of whole-word embedding cannot effectively capture high-order collaborative knowledge and generalize well to unseen users/items for recommendation.
In addition, due to the nature of discrete tokens in language models, using continuous indexing makes it challenging to align LLMs in recommender systems closely.

To address such challenges, we propose a novel LLM-based framework for recommender systems (\textbf{\ourname{}}), in which a novel tokenization strategy is proposed to tokenize numerical ID (i.e., identifiers) of users and items by seamlessly integrating high-order collaborative knowledge into LLMs. 
Meanwhile, a generative retrieval paradigm is developed to generate item representations and retrieve appropriate items for collaborative recommendations.
Our major contributions are summarized as follows:
\begin{itemize}[leftmargin=*]
    \item We introduce a principle strategy named \textbf{Masked Vector-Quantized Tokenizer} to tokenize users and items tailored to LLMs, which contributes to incorporating high-order collaborative knowledge in LLM-based recommendations.
    More specifically, two novel mechanisms (i.e., masking and $K$-way encoder) are designed to enhance the generalization capability of the proposed tokenization method in LLM-based recommendations.

    \item We propose a novel framework (\textbf{TokenRec}) for recommender systems in the era of LLMs, where a generative retrieval paradigm is designed to effectively and efficiently recommend top-$K$ items for users rather than directly generating tokens in natural language. 

    \item We conduct extensive experiments on four widely used real-world datasets to empirically demonstrate the effectiveness of our proposed \ourname{}, including the superior recommendation performance and its generalization ability in predicting new and unseen users' preferences. 
\end{itemize}

The remainder of this paper is structured as follows. 
Section \ref{sec:methodlogy} introduces the
proposed approach, which is evaluated in Section \ref{sec:Experiments}. 
Then, Section \ref{sec:relatedwork} summarizes the recent development of collaborative filtering and LLM-based RecSys. Finally, conclusions are drawn in Section \ref{sec:conclusion}.

\section{The Proposed Method}
\label{sec:methodlogy}
This section will begin by reformulating the collaborative recommendation as a language-processing task.
Then, we provide an overview of the proposed \ourname{}, followed by a detailed explanation of each model component.
Finally, we will discuss the model training and inference of \ourname{}.

\subsection{Notations and Definitions} \label{sec:task}
Let $\mathcal{U}=\{u_1,u_2,...,u_n\}$ and $\mathcal{V}=\{v_1,v_2,...,v_m\}$ be the sets of users and items, respectively, where $n$ is the number of users, and $m$ is the number of items.
Moreover, we use $\mathcal{N}_{(u_i)}$ to denote the item set that user $u_i$ has interacted in the history. 
As the traditional collaborative filtering methods, user $u_i$ and item $v_j$ can be embedded into low-dimensional latent vectors (i.e., collaborative representations of users and items)  $\mathbf{p}_i \in \mathbb{R}^d$ and $\mathbf{q}_j \in \mathbb{R}^d$ respectively, where $d$ is the length of the vector.

In general, the goal of a recommender system is to understand users’ preferences by modeling interactions (e.g., clicks and bought) between users $\mathcal{U}=\{u_1,u_2,...,u_n\}$ and items $\mathcal{V}=\{v_1,v_2,...,v_m\}$.
As a widely used solution, collaborative filtering (CF) techniques are developed to learn user and item representations from historical user-item interactions. 
Thus, we reformulate the CF recommendation in the language model paradigm.
Assume that we only have token IDs  $\mathcal{T}_i$ and $\mathcal{T}_j$ for each user $u_i$ and item $v_j$.
By integrating these IDs into textual prompts $\mathcal{P}$, LLM generates the representation  $\mathbf{z}_i \in \mathbb{R}^d$ of items that a user $u_i$ may like, expressed as:
\begin{align}
    \label{eq:task_predict}
    \mathbf{z}_i = {\rm LLM}(\mathcal{P}, \mathcal{T}_i, \{\mathcal{T}_j | v_j \in \mathcal{N}_{(u_i)}\}).
\end{align}
Notably, the interacted items $\mathcal{N}_{(u_i)}$ will be placed into the language model in a \emph{non-sequential way} to accommodate the setting of collaborative filtering.

\subsection{An Overview of the Proposed Framework}
\begin{figure*}[t]
\centering
{\includegraphics[width=\textwidth]{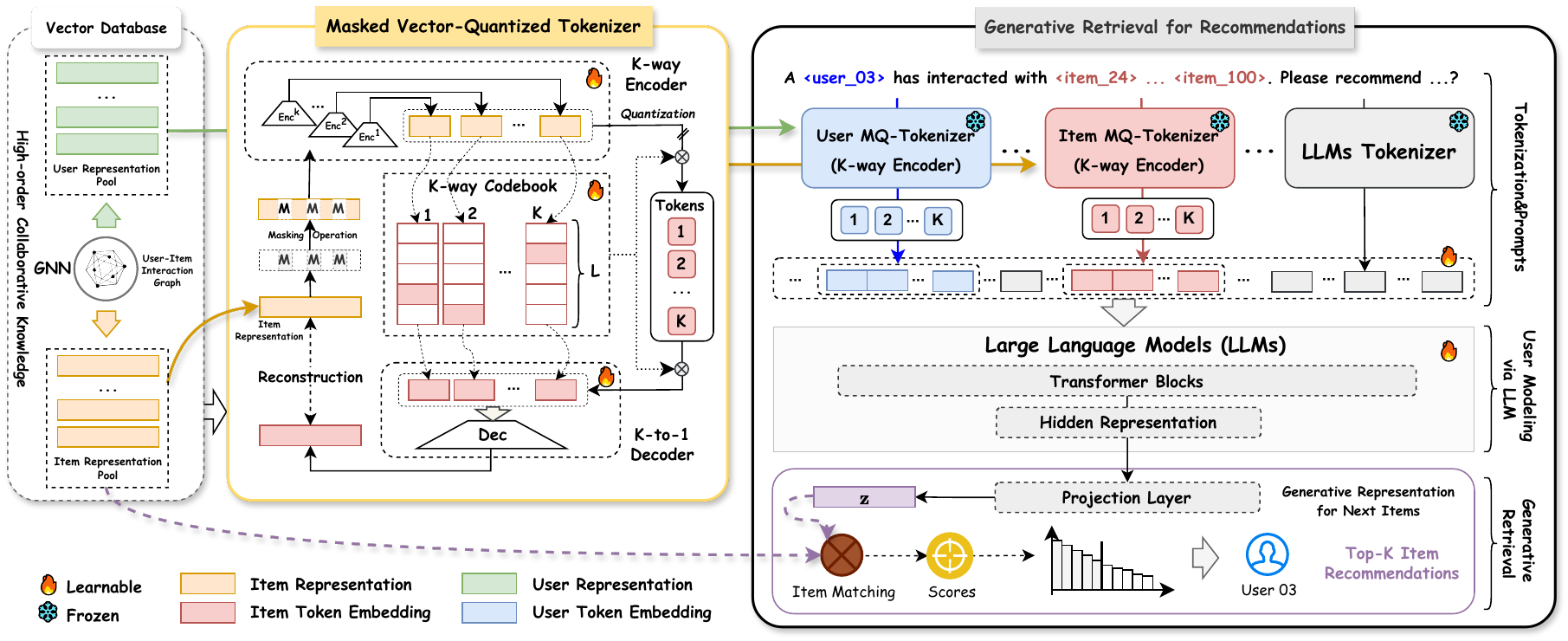}}
\vskip -0.1in
\caption{The overall framework of the proposed \ourname{}, 
which consists of the masked vector-quantized tokenizer with a $K$-way encoder for item ID tokenization and the generative retrieval paradigm for recommendation generation. Note that we detail the item MQ-Tokenizer while omitting the user MQ-Tokenizer for simplicity.} 
\label{fig:method}
\vskip -0.1in
\end{figure*}

To better align natural language with recommendation tasks, we propose a novel LLM-based generative recommendation framework (TokenRec).
As shown in Figure~\ref{fig:method}, the proposed framework consists of two key modules, namely \textbf{Masked Vector-Quantized (MQ) Tokenizer for Users and Items} and \textbf{Generative Retrieval for Recommendations}.
The first module aims to address the fundamental task of ID tokenization in LLM-based recommendations so as to seamlessly integrate users\&items (i.e., numeric ID) into natural language form. 
However, tokenizing users\&items faces tremendous challenges due to the huge number of users and items in recommender systems.
To address the emerging challenges, we introduce a Masked Vector-Quantized Tokenizer (MQ-Tokenizer) to learn specific codebooks and represent users\&items with a list of special tokens through encoder and decoder networks.
The goal of the second module is to perform user modeling via LLM for personalized recommendations.
To achieve this, a generative retrieval paradigm is introduced to retrieve the $K$-nearest items from the whole item set for generating a personalized top-$K$ recommendation list in an effective and efficient manner.
The details of the proposed method, TokenRec, will be described in the following sections.

\newcommand{\Ein}{{\vE_{\text{in}}}}

\subsection{Masked Vector-Quantized Tokenizers for Users and Items}
Instead of assigning each user and item with a specific token (causing a significant increase in vocabulary size), we propose a novel strategy to tokenize users and items to align with natural language, where quantization techniques are developed to represent each user and item with certain discrete indices (i.e., tokens).  
More specifically, as one of the most representative quantization techniques, Vector Quantization (VQ) aims to convert inputs into a set of discrete tokens (i.e., codes) and learn a discrete latent representation space (i.e., codebook) by reconstructing the original inputs~\cite{esser2021taming,ramesh2021zero,van2017neural}. 
Moreover, in order to capture \emph{high-order collaborative knowledge} from users' interaction history, we propose to conduct the vector quantization on the well-trained representations learned from advanced GNN-based recommendations. 
However, naively applying vector quantization to index/tokenize users and items is fraught with critical challenges due to poor generalization capability. 
For example, certain noise and propagation errors in the cascade process of user and item tokenizations in the LLMs greatly hinder the expressiveness of the vanilla VQ. 

To tackle these challenges, we propose a Masked Vector-Quantized Tokenizer (\textbf{\ourvqname{}}) for tokenizing users and item IDs, where two novel strategies, namely masking and $K$-way encoding, are designed to enhance the generalization capability of our proposed tokenization in LLM-based recommendations. 
More specifically, the proposed \ourvqname{} consists of a masking operation on the inputs (i.e., users\&items representations), a $K$-way encoder for multi-head feature extraction with a corresponding $K$-way codebook for latent feature quantization, and a $K$-to-1 decoder that reconstructs the input representations with the quantized features.
It is worth mentioning that the proposed tokenization method will design the specific MQ-Tokenizers for the user and item respectively (i.e., \emph{Item MQ-Tokenizer} and \emph{User MQ-Tokenizer}), in which they share the same architecture. 
Thus, for simplicity, we will detail the item MQ-Tokenizer for item ID tokenization while omitting the user MQ-Tokenizer, as shown in Figure~\ref{fig:method}.

\subsubsection{\textbf{Collaborative Knowledge}}
The primary objective of our proposed MQ-Tokenizer lies in capturing high-order collaborative knowledge into latent representations through vector quantization.
Collaborative knowledge has been widely proven informative for predicting user preferences in recommender systems~\cite{wang2019neural,fan2019graph}, as it reveals the in-depth behavioral similarity among users (or items). 
However, it is challenging for LLMs to map textual descriptions of the high-order collaborative relationships between users and items explicitly with natural language~\cite{wang2024can}. 
Meanwhile, as the most advanced collaborative filtering techniques, graph neural networks (GNNs) are proposed to learn user and item representations by capturing high-order collaborative signals among users and items in recommendations~\cite{he2020lightgcn,fan2022graph}. 
Despite great potential, it is challenging to fully harness the high-order signals from user-item interactions in natural language for LLM-based recommendations. 
Thus, we propose to perform vector quantization on the collaborative representations learned by advanced GNN-based methods for indexing/tokenizing users and items. 
Specifically, our \ourvqname{} assigns each user and item a small number of discrete tokens generated by quantizing the GNN-based representations.
In other words, it is suggested that users and items in close proximity to each other in the latent space of collaborative knowledge (i.e., user/item representations) are more likely to share similar tokens/indices, which naturally aligns LLMs with recommendations by tokenizing users and items in natural language.

\subsubsection{\textbf{Masking Operation}}
On the basis of introducing the collaborative representations learned from GNN-based collaborative filtering methods, which are stored in a vector database, we propose to randomly mask these representations to create a challenging task that enables the tokenizer to build a comprehensive understanding and generalize well.
Specifically, we introduce an element-wise masking strategy $\mathcal{E}$ following Bernoulli distribution as follows:
\begin{align}
    \label{eq:bernoulli}
    \mathcal{E} \sim {\rm Bernoulli}(\rho),
\end{align}
where $\rho$ is the masking ratio. 
The Bernoulli distribution is the discrete probability distribution of a random variable, which takes the value 1 with probability $\rho$ and the value 0 with probability $1-\rho$.
Given the collaborative representations $\mathbf{p}_i$ and $\mathbf{q}_j$, the masking process can be presented by
\begin{align}
    \label{eq:masking}
    \mathbf{p}'_i = \text{Mask}(\mathbf{p}_i, \mathcal{E}), \ \mathbf{q}'_j = \text{Mask}(\mathbf{q}_j, \mathcal{E}),
\end{align}
where $\mathbf{p}'_i$ and $\mathbf{q}'_j$ are the masked representations of user $u_i$ and item $v_j$, respectively.
It should be noted that $\mathcal{E}$ randomly generates masks at each training epoch to create multifarious samples so as to enhance the generalization capability in the proposed tokenizers.

\subsubsection{\textbf{$K$-way Encoder and Codebook}}
In light of the masked collaborative representations of items, we propose a novel vector quantization framework to tokenize each item as a handful of discrete tokens.
To be more specific, a $K$-way encoder $\text{Enc}(\cdot)$ is developed to learn a corresponding $K$-way codebook $\mathbf{C} = \{ \mathbf{c}^1, \mathbf{c}^2, ..., \mathbf{c}^k,..., \mathbf{c}^K \}$ for items, where $\mathbf{c}^k \in \mathbb{R}^{L \times d_c}$ is a latent space (i.e., the $k$-th sub-codebook) with $L$ pairs of codeword (i.e., token\footnote{In this work, the terms 'codeword' and 'token' are used interchangeably. 'Codeword' is a common term in the context of codebooks, while 'token' is more widely used in the language modeling literature. }) and  $w^k$ and its $d_c$-dimensional codeword embedding $\mathbf{c}^k_{w^k}$. 
The idea behind this is to represent the users/items with entries of a learned codebook (indexed by discrete tokens),  providing great potential to tokenize the discrete users and items based on their collaborative representations in an LLM-compatible manner.

In general, the proposed quantization method for user/item tokenization involves two main steps.
Taking item $v_j$ as an example, the $K$-way encoder first employs $K$ different encoders $\text{Enc}^k(\cdot)$ to encode the masked item representation $\mathbf{q}'_j$ and generate  $K$ corresponding latent vectors $\{ \mathbf{a}^k_j \}^K_{k=1}$  as follows: 
\begin{align}
\label{eq:vq-mlp}
    \mathbf{a}^k_j = \text{Enc}^k(\mathbf{q}'_j) =\text{MLP}^k(\mathbf{q}'_j), 
\end{align}
where $\mathbf{a}^k_j \in \mathbb{R}^{d_c}$, and each encoder $\text{Enc}^k(\cdot)$ can be implemented as a multilayer perceptron network (MLP) with three hidden layers. 
Powered by different encoders, the proposed $K$-way Encoder enables multiple attentions and uncovers different patterns on the inputs to enhance the generalization capability towards users\&items representations.

The next step is to quantize the encoded vectors $\{ \mathbf{a}^k_j \}^K_{k=1}$  into discrete tokens (i.e., indices) by looking up nearest neighbors in the learnable $K$-way codebook $\mathbf{C}=\{\mathbf{c}^1, \mathbf{c}^2, ...,\mathbf{c}^k,..., \mathbf{c}^K\}$.
More specifically, given the item $v_j$ and a sub-codebook $\mathbf{c}^k \in \mathbb{R}^{L \times d_c}$ from the $k$-th encoder $\text{Enc}^k$, 
a practical approach is leveraging the Euclidean distance to calculate similarity scores between the encoded vector $\mathbf{a}^k_j$ and all the codeword embeddings $\{ \mathbf{c}^k_l \}^L_{l=1}$, so as to  find the nearest embedding (i.e., token embedding) that can be used to effectively represent the encoded vector $\mathbf{a}^k_j$: 
\begin{align}
\label{eq:quantization}
    w^k_j = \text{arg min}_l \Vert \mathbf{a}^k_j - \mathbf{c}_l^k \Vert^2, \\
    \text{Quantize}(\mathbf{a}^k_j) = \mathbf{c}^k_{w^k_j},
\end{align}
where $w^k_j$ denotes the codeword (i.e., ID token) of the nearest neighbor at sub-codebook $\mathbf{c}^k$ for the item $v_j$, and $\Vert \cdot \Vert^2$  denotes the $l_2$ norm of the variables.

In other words, the proposed Item MQ-Tokenizer can  tokenize the discrete ID  of item $v_j$ to $K$ discrete codebook tokens along with their corresponding codeword embeddings in LLM-based recommender systems as follows: 
\begin{align}
\text{item} ~v_j &\rightarrow  \text{tokens: }  \{w_j^1, w_j^2, ..., w_j^K \} \\
& \rightarrow  \text{tokens' embeddings: } [\mathbf{c}^1_{w^1_j},  \mathbf{c}^2_{w^2_j},  ..., \mathbf{c}^K_{w^K_j} ].
\label{eq:selected_codeword}
\end{align}
It is worth noting that we can also apply a similar quantification process to tokenize users in LLM-based recommendations.

\subsubsection{\textbf{$K$-to-1 Decoder}}
After the $K$-way encoder, a $K$-to-1 decoder $\text{Dec}(\cdot)$ is introduced to conduct input reconstruction.
The basic idea is that $K$ different embeddings indexed by $K$ discrete tokens in the $K$-way codebook $\mathbf{C}$ are fed into the $K$-to-1 decoder to reconstruct the input representations of user $\mathbf{p}_i$ or item $\mathbf{q}_j$. 
Mathematically, given item $v_j$ and its quantized tokens $\{w_j^1, w_j^2, ..., w_j^K \}$, 
the decoder first performs average pooling and then generates the reconstructed input representation $ \mathbf{r}_j$ via a three-layer MLP as follows: 
\begin{align}
   \label{eq:mvq_decoding}
   \mathbf{r}_j = \text{Dec}( \{w_j^1, w_j^2, ..., w_j^K \} ) = \text{MLP}(\frac{1}{K} \sum_{k=1}^K \mathbf{c}^k_{w^k_j}). 
\end{align}

\subsubsection{\textbf{Learning Objective}}
In order to effectively learn the $K$-way encoder, codebook, and $K$-to-1 Decoder for user and item MQ-Tokenizers, a \emph{reconstruction loss} is designed to reconstruct the prototype inputs from the set of discrete vectors.
To be more specific, given the item $v_j$, the reconstruction loss encourages the reconstructed representation $\mathbf{r}_j$ from the $K$-to-1 decoder to approximate original item representation $\mathbf{q}_j$ learned from GNNs, which can be defined as:
\begin{align}
\label{eq:mq_recon}
    \mathcal{L}^{Item}_{recon} = \Vert \mathbf{q}_j - \mathbf{r}_j \Vert^2.
\end{align}
However, the \emph{arg~min} operation in Eq.~\eqref{eq:quantization} is non-differentiable~\cite{berthet2020learning,lorberbom2019direct}, leading to an intractable computation for the gradient back-propagation in the reconstruction loss $\mathcal{L}^{Item}_{recon}$ during optimization. 
To this end, a straight-through gradient estimator~\cite{bengio2013estimating,van2017neural} is introduced to directly assign the gradients of the decoder inputs (i.e., the selected tokens' embeddings) to the encoder outputs (i.e ., the encoded representations), so as to optimize the encoders and decoder. 

Meanwhile, optimizing the reconstruction loss $\mathcal{L}^{Item}_{recon}$ cannot provide any gradients to update items' $K$-way codebook $\mathbf{C}$. To address this, a \emph{codebook loss} $\mathcal{L}^{Item}_{cb}$  for items is further designed to bring the selected  token's embedding $\mathbf{c}^k_{w^k} $ close to the outputs of the $K$-way encoder $\text{Enc}^k(\cdot)$ using the $l_2$ error for updating $K$-way codebook $\mathbf{C}$ as follows:
\begin{align}
\label{eq:mvq_codebook}
\mathcal{L}^{Item}_{cb} &= \sum_{k=1}^K \Vert \text{sg}[~\text{Enc}^k(\mathbf{q}'_j)~] - \mathbf{c}^k_{w^k_j} \Vert^2,
\end{align}
where $\text{sg}[\cdot]$ denotes the a stop-gradient operator.  
To be specific, the gradient of the variable in $\text{sg}[\cdot]$ takes the value 0 when performing back-propagated gradient calculation.

Furthermore, to encourage a smooth gradient passing for the \emph{arg~min}  operation from Eq.~\eqref{eq:quantization}, a \emph{commitment loss} $\mathcal{L}^{Item}_{cm}$ is introduced to prevent the encoded features from fluctuating too frequently from one codeword to another. Unlike the codebook loss, the commitment loss only applies to the encoder weights, which can be calculated by: 
\begin{align}
\label{eq:mvq_codebook}
\mathcal{L}^{Item}_{cm}
    = \sum_{k=1}^K \Vert ~\text{Enc}^k(\mathbf{q}'_j)~ - \text{sg}[\mathbf{c}^k_{w^k_j}] \Vert^2.
\end{align}

Finally, the overall optimization objective for item MQ-Tokenizer can be formulated by incorporating the aforementioned losses to jointly update the $K$-way encoder, the $K$-way codebook, and the $K$-to-1 decoder:
\begin{align}
\label{eq:item_overall}
    \mathcal{L}^{Item}_{MQ} &= \mathcal{L}^{Item}_{recon} + \mathcal{L}^{Item}_{cb} + \beta^{Item} \times \mathcal{L}^{Item}_{cm}, 
\end{align}
where $\beta^{Item}$ is a hyper-parameter that aims to balance the importance of the commitment loss in the overall objective.
Additionally, the aforementioned optimization objective for optimizing the user MQ-Tokenizer can be designed as follows:
\begin{align} 
\label{eq:user_overall}
    \mathcal{L}^{User}_{MQ} &= \mathcal{L}^{User}_{recon} + \mathcal{L}^{User}_{cb} + \beta^{User} \times \mathcal{L}^{User}_{cm}.
\end{align}

\subsection{Generative Retrieval for Recommendations}
In this subsection, we introduce a novel framework that takes advantage of LLMs for recommendations.
Particularly, a generative retrieval paradigm is designed to generate item representations and retrieve appropriate items for collaborative recommendations.
To align large language models with collaborative recommendations, the proposed framework involves three key components to reformulate collaborative filtering: tokenization\&prompting, user modeling via LLM, and generative retrieval.

\subsubsection{\textbf{Tokenization\&Prompts}}
Tokenization is the most crucial process of splitting the textual input and output into smaller units that language models can process.
In general, LLMs have thousands of tokens in their vocabularies (i.e., \emph{in-vocabulary words}).
For instance, a representative LLM, LLaMA~\cite{touvron2023llama}, has a vocabulary size of 32,000.
Nonetheless, the number of tokens in LLMs remains relatively small compared to the enormous number of users and items in real-world recommendation scenarios, often reaching millions or billions.
Therefore, we introduce \emph{out-of-vocabulary tokens} learned from the proposed \emph{user and item MQ-Tokenizers} to facilitate the tokenization of user and item IDs.
Using the \ourvqname{}s, only a limited number of $K \times L$ out-of-vocabulary (OOV) tokens are required to tokenize millions or billions of users or items effectively. 
For example, in our experiments (see Section~\ref{sec:Experiments}), we can use only 1,536 (i.e., 3 $\times$ 512) OOV tokens to tokenize a total of 39,387 items in the Amazon-Clothing dataset. 
The vocabulary expansion for tokenizing users and items by \ourvqname{}s is much more efficient and affordable for LLM-based recommendations. 
In other words, textual contents are tokenized by the LLM tokenizer (e.g., SentencePiece), while user and item IDs are tokenized into $K$ discrete OOV tokens via the corresponding MQ-Tokenizers, respectively.

To further enhance the capabilities of LLMs, prompting has been developed to provide explicit guidance to LLMs, enabling them to make better predictions for downstream tasks~\cite{zhao2023survey,zhao2024recommender}. 
Building upon this insight, we design several prompts instructing the LLM backbone to understand users' preferences for making recommendations.
In these prompts, the ID tokens provided by the well-established \ourvqname{}s are used to represent users and items in the language space of LLM-based recommender systems. 
For example, given the sub-codebook number $K=3$, two representative prompts in our method and their user\&item tokenizations are defined as follows:
\begin{tcolorbox}
\label{box:prompt1}
Prompt 1 (without user's historical interactions): \\ I wonder what the \colorbox{cyan!30}{\textbf{user\_03}} will like. Can you help me decide?
\tcblower
$\Longrightarrow$ I wonder what the \\ \colorbox{cyan!30}{$user\_\left \langle u1\text{-}128 \right \rangle \left \langle u2\text{-}21 \right \rangle\left \langle u3\text{-}35 \right \rangle $}
will like. Can you help me decide? 

\end{tcolorbox}

\begin{tcolorbox} 
Prompt 2 (with user's historical interactions):  \\
According to what items the \colorbox{cyan!30}{\textbf{user\_03}} has interacted with: \colorbox{red!30}{\textbf{item\_08}}, \colorbox{red!30}{\textbf{item\_24}}, \colorbox{red!30}{\textbf{item\_63}}. Can you describe the user's preferences? 
\tcblower
$\Longrightarrow$ According to what items the

\colorbox{cyan!30}{$user\_\left \langle u1\text{-}128 \right \rangle \left \langle u2\text{-}21 \right \rangle\left \langle u3\text{-}35 \right \rangle $} has interacted with:

\colorbox{red!30}{$item\_\left \langle v1\text{-}42 \right \rangle \left \langle v2\text{-}12 \right \rangle\left \langle v3\text{-}98 \right \rangle $}, 
\colorbox{red!30}{$item\_\left \langle v1\text{-}42 \right \rangle \left \langle v2\text{-}12 \right \rangle\left \langle v3\text{-}87 \right \rangle $},
\colorbox{red!30}{$item\_\left \langle v1\text{-}42 \right \rangle \left \langle v2\text{-}53 \right \rangle\left \langle v3\text{-}128 \right \rangle $}. 

Can you describe the user's preferences?
\end{tcolorbox}

Technically, Prompt 1 and Prompt 2 showcase the input prompts without and with item interactions as supporting information, respectively, in which $\left \langle uk\text{-}\cdot \right \rangle$ denotes the out-of-vocabulary (OOV) tokens in the $k_{th}$ sub-codebook.
For instance,  $\left \langle u2\text{-}21 \right \rangle$ represents the $21_{st}$ token in the second sub-codebook for the \textbf{user\_03}.
This also applies to tokenizing items in LLMs.

\subsubsection{\textbf{User Modeling via LLM}}
The goal of the user modeling component is to capture users' preferences for generating the representations of items that user $u_i$ may like.
A typical input $\mathcal{X}_i$ of our LLM backbone can be formed by selecting a prompt template $\mathcal{P}$ and the corresponding ID tokens for user $u_i$ and his/her interacted items $\mathcal{N}_{(u_i)}$ in the history as follows: 
\begin{align}
\label{eq:llm_input}
    \mathcal{X}_i \rightarrow (\mathcal{P}, \mathcal{T}^c_{u_i}) \ or \ (\mathcal{P}, \mathcal{T}^c_{u_i}, \{ \mathcal{T}^c_{v_j} | v_j \in \mathcal{N}_{(u_i)}\}),
\end{align}
where $\mathcal{T}^c_{u_i}$ represents the ID tokens generated by the user \ourvqname{} for user $u_i$. $\{ \mathcal{T}^c_{v_j} | v_j \in \mathcal{N}_{(u_i)} \}$ denotes the ID tokens generated by our item \ourvqname{} for the items that user $u_i$ has interacted with. 
It is worth noting that we can randomly shuffle their interactions in $\mathcal{N}_{(u_i)}$ for neglecting the consideration of user $u_i$'s sequential signals towards items.

Under the conventional text-to-text generation paradigm (e.g., P5), the user modeling process involves receiving a text input $\mathcal{X}_i$ for user $u_i$ and generating descriptive texts for potential items in \textbf{an auto-regressive manner}, which can be formalized as:
\begin{align}
\label{eq:hidden_states}
    \mathcal{T}_t = \text{LLM}(\mathcal{X}_i, \mathcal{T}_{i:t-1}), 
\end{align}
where $\mathcal{T}_t$ represents the token being generated at the $t_{th}$ position, while $\mathcal{T}_{1:t-1}$ denotes the previously generated tokens from the LLM.
In contrast, our user modeling process differs from the previous approach.
To be more specific, we take the input $\mathcal{X}_i$ and pass it through the LLM backbone, denoted as $\text{LLM4Rec}(\cdot)$ to generate a hidden representation  $\mathbf{h}_i$ that reflects the model's comprehension of user $u_i$'s preferences for next-items recommendations. 
Mathematically, we can express this process as follows: 
\begin{align}
\label{eq:hidden_states}
    \mathbf{h}_i = \text{LLM4Rec}(\mathcal{X}_i).  
\end{align}
In other words, the final encoded representation $\mathbf{h}_i$ can be considered as user $u_i$'s generative preferences of the next items for making personalized recommendations.
{Within our framework, the LLM backbone functions as a powerful query encoder, positioned to excel over traditional deep neural networks in user modeling across several critical dimensions: (a) comprehending personalized user queries for recommendations through diverse prompts with/without user’s historical interactions; (b) interpreting users' preferences leveraging LLM's impressive abilities in reasoning; and (c) generating desired outcomes: LLMs can beyond text~\cite{li2023mage,zhu2024beyond}.}

\subsubsection{\textbf{Generative Retrieval}}
In general, LLM-based recommender systems employ auto-regressive generation to decode recommendations in natural language~\cite{geng2022recommendation,zheng2023adapting}, such as producing textual strings like "\emph{item\_1234}" or \emph{"the user love electronics ...}". 
However, the beam search decoding in LLMs can be very time-consuming during auto-regressive generation~\cite{wang2024rethinking}, which is impractical for various real-time recommendation scenarios. 
Moreover, due to the hallucination issue, generating accurate item titles and descriptions is highly challenging when making personalized recommendations. 
For example, items' title ``\emph{iPhone SE, 256 GB, starlight}'' and ``\emph{iPhone 15, 256 GB, starlight}'' share most tokens but are significantly different products - with ``\emph{iPhone 15, 256 GB, starlight}' being a non-factual product. 
The hallucination issue is likely more severe in e-commerce platforms, where billions of products are sold, leading to invalid item identifiers for recommendations. 
Furthermore, title generation in LLM-based recommendations cannot generate unseen items in the fine-tuning stage, which is often infeasible in practice. 
To this end, we propose a generative retrieval paradigm for LLM-based recommendations, where a simple but effective and efficient strategy is designed to project generative users' preferences for retrieving potential items from the whole item pool. 

More specifically, the hidden state $\mathbf{h}_i$ from $\text{LLM4Rec}(\cdot)$ will be projected to a latent representation $\mathbf{z}_i \in \mathbb{R}^d$ through a projection layer $\text{Proj}(\cdot)$, to make the alignment between the LLM-generated representation and item representations learned from GNNs as follows:
\begin{align}
\label{eq:hidden_states}
    \mathbf{z}_i = \text{Proj}(\mathbf{h}_i),
\end{align}
where $\text{Proj}(\cdot)$ can be modeled by a three-layer MLPs. 
Note that $\mathbf{z}_i $ can be considered the generative representation of the next recommended items for user $u_i$'. 
After that, we propose to retrieve the $K$-nearest items from the whole item set $\mathcal{V}$ for generating the personalized top-$K$ recommendation list.
This can be achieved by measuring the similarity scores between the target user's generative preference and high-order items' representations from the well-trained GNN-based recommendation method.
For example, the predicted similarity score $y_{ij}$ of user $u_i$ towards item $v_j$ can be calculated by a matching function (e.g., cosine similarity) between user $u_i$'s generative item representation $\mathbf{z}_i$ and GNN-based item's $v_j$ representation $\mathbf{q}_j$ which is stored in a vector database: 
\begin{align}
\label{eq:scoring}
    y_{ij} = \frac{\mathbf{z}_i \mathbf{q}_j}{\Vert \mathbf{z}_i \Vert \Vert \mathbf{q}_j \Vert}.
\end{align}

{To sum up, given the scores of all items in the item base, the proposed \ourname{} can easily retrieve top-$K$ items for users rather than directly generating tokens in natural language. This approach offers advantages in terms of efficiency in inference and avoiding hallucinations~\cite{wang2024rethinking}, distinguishing \ourname{} from the majority of existing language model-based Recommender Systems that struggle with their time-consuming auto-regressive decoding and beam search processes.
Furthermore, unseen items in the fine-tuning stage can be retrieved by only updating the item representations pool $\mathcal{V}$ instead of retraining the entire model. Lastly, this two-tower-like structure~\cite{yang2020mixed} shows promise in facilitating seamless alignment between textual information on the query side and collaborative knowledge on the hidden representation side.}

\subsection{TokenRec's Training and Inference }
\subsubsection{\textbf{Training}}Technically, the proposed \ourvqname{} is responsible for incorporating high-order collaborative knowledge into ID tokenization, while the LLM4Rec backbone is used to capture user preferences and generate the list of items for recommendations.
One straightforward approach is to update these two components jointly.
However, the large gap between quantization and language processing makes it difficult to achieve updates synchronously.
Here, we first train \ourvqname{}s for users/items and then freeze the well-trained \ourvqname{}s components to guide the tuning process of our LLM backbone.
More specifically, our training process is as follows:
\begin{itemize}[leftmargin=*]
    \item \textbf{Step 1. Training Users\&Items \ourvqname{}s}.
In order to learn users\&items ID tokenization, our initial focus lies in training \ourvqname{}s to quantize the collaborative representations for users and items independently. 
Specifically, we use the combined losses as given in Eq. \eqref{eq:item_overall} and Eq. \eqref{eq:user_overall} to train the \ourvqname{}s for items $\mathcal{V}$ and users $\mathcal{U}$, separately.

\item  
\textbf{Step 2. Tuning the LLM4Rec for Generative Retrieval}. 
In this step, we tune the LLM backbone (e.g., T5), LLM token embeddings, and the projection layer for generative retrieval recommendations while keeping the \ourvqname{}s frozen. 
More specifically, the objective of generative retrieval is to identify potential items from the whole item pool,
such that top-$K$ ranked items are relevant to the specific query (i.e., generative item representations of users).
To achieve such an objective, a general solution is to perform metric learning by predicting the relative similarity or distance based on the dense representations between inputs~\cite{cakir2019deep,mcfee2010metric,liu2009learning}. 
In other words, the proposed TokenRec aims to perform the nearest neighbor retrieval, achieved by calculating ranking scores between the user's generative item representation and collaborative item representations learned from GNNs, so as to retrieve top-$K$ next items to the target user for personalized recommendations.
Mathematically, we utilize a pairwise ranking loss for our tuning process, which encourages the query closer to the positive set (i.e., when $\lambda=1$) than to the negative set by a fixed margin $\gamma$ (i.e., when $\lambda=-1$)  as follows:
\begin{align}
\mathcal{L}_{\text{LLM4Rec}} = 
\begin{cases}
1 - \text{sim} (\mathbf{z}_i, \mathbf{q}_j),  & \text{if} \ \lambda=1 \\
\max (0, \text{sim} (\mathbf{z}_i, \mathbf{q}_j) - \gamma), & \text{if} \ \lambda=-1
\end{cases}
\label{eq:llmloss}
\end{align}
where $\mathbf{z}_i$ and $\mathbf{q}_j$ denote the generative item representation of user $u_i$ and the collaborative representation of item $v_j$, respectively. 
$\text{sim}(\cdot, \cdot)$ is a metric function to measure the similarity between dense representations, such as cosine similarity. 
$\lambda$ indicates whether user $u_i$ has interacted with item $v_j$.
Moreover, $\gamma$ is the margin value for negative pairs.
It ensures that when the representations of a negative pair are already adequately distant, there is no need to expend additional effort in increasing the distance between them.
This mechanism allows more focused training on pairs that are more challenging for recommendations. 
More specifically, we use cosine similarity as the metric function to maximize the similarity between user $u_i$' s generative representation $\mathbf{z}_i$  of next item and the \emph{positive item $v_j$ representation}  $\mathbf{q}_j$ (i.e., item $v_j$ in user $u_i$'s historical interactions) when  $\lambda=1$, while $\lambda=-1$ otherwise.
\end{itemize}

\begin{figure}[b]
\centering
{\includegraphics[width=0.99\linewidth]{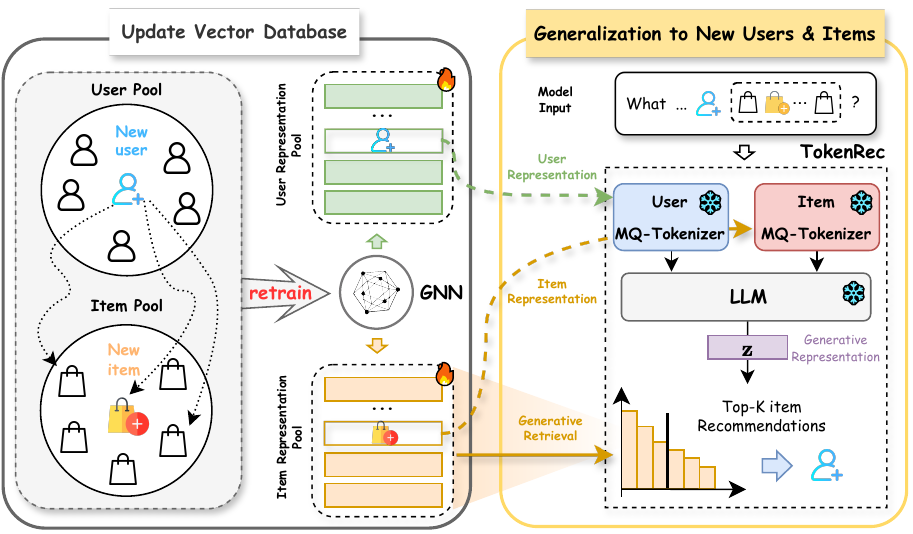}}
\vskip -0.05in
\caption{The TokenRec's efficiency and generalization capability for new users and items during the inference stage.
Rather than retraining the MQ-Tokenizers and LLM backbone, which can be computationally expensive and time-consuming, only the GNN needs to be updated for learning representations for new users and items.}
\label{fig:inference}
\end{figure}

\subsubsection{\textbf{Inference}}
The typical inference of LLMs refers to the process of generating target tokens.
It suffers from laborious generation and insufficient modeling for unseen users and items.
By utilizing the generative retrieval framework, our inference process overcomes these challenges in an efficient way.
Specifically, the advantages of our proposed method can be summarized as follows:
\begin{itemize}[leftmargin=*]
    \item \textbf{Efficient Recommendations}. \ourname{} proposes a novel LLM-empowered collaborative recommendation framework in generative retrieval paradigms, bypassing the time-consuming decoding process. 
    In particular, the proposed LLM-based recommendation framework aims to output generative item representation and retrieve appropriate items for collaborative recommendations instead of a sequence of discrete tokens (e.g., item titles). This contributes to reducing the considerable computing cost of online RecSys. The efficiency evaluation can be found in Section~\ref{subsec:Efficiency} and Table~\ref{tab:efficient}.

    \item \textbf{Generalizability to New Users and Items}. 
    The proposed architecture can provide robust ID tokenization for unseen users and items without fine-tuning the LLM4Rec component.
    {In practice, web platforms often encounter a frequent influx of new products (items) and users. These entities might have a few interactions. For instance, in social media platforms, numerous new users join daily, sharing tweets on diverse topics, following others, and interacting through likes, retweets, and replies. However, these interactions are not encompassed within the training corpus of recommendation models, rendering them as unseen users/items. Maintaining regular updates to the foundational model for this daily influx proves to be a costly endeavor for existing LLM-based RecSys. To address this problem, we propose to leverage external lightweight GNN models to learn corresponding representations of these unseen items/users, so that TokenRec can perform inference on them without requiring retraining of LLM backbone and MQ-Tokenizers. As shown in Figure~\ref{fig:inference}, when new users and items are added to the system, retraining is only required for the collaborative filtering component (e.g., MF and LightGCN) to capture collaborative knowledge in learning representations of these users and items (i.e., updating vector database). 
    In contrast, the MQ-Tokenizers and LLM backbone can remain frozen and perform well, thanks to the masking and $K$-way encoder mechanisms in vector quantization.
    This addresses the cold-start problem for LLM-based RecSys, eliminates the need for retraining LLM-related components, and therefore saves huge computational resources: compared to the finetuning of LLMs, the training of GNNs is far more efficient~\cite{zhang2024ltgnn}.}
    Notably, the capability to adapt to new users and items is demonstrated in Section~\ref{subsec:Generalizability} and Table~\ref{tab:generalizability}.

    \item 
    \textbf{Concise Prompts}. \ourname{} provides an inference alternative that relies solely on user ID tokens, e.g., Prompt 1 in Section \ref{box:prompt1}, for LLM-based recommendation generation.
    This is made possible by incorporating the collaborative knowledge of users into user ID tokens through our \ourvqname{}.
    By doing so, \ourname{} eliminates the necessity of including interacted items in inputs, thus reducing significant computing resources during inference.
    Additionally, this mechanism proves advantageous when dealing with users who have interacted with a large number of items, effectively avoiding the prevalent issue of context length limitation of many LLMs \cite{liu2024lost,kaneko2023reducing}, e.g., 512 tokens in T5~\cite{raffel2020t5} and 2048 tokens in ChatGPT.
    The recommendation performance using user ID tokens only for model input can be observed in Table~\ref{tab:comparison_LastFM&ML1M} and Table~\ref{tab:comparison_Beauty&Clothing}.
\end{itemize}
\section{Experiment}
\label{sec:Experiments}

\subsection{Experimental Settings}

\subsubsection{\textbf{Datasets}}
To demonstrate the effectiveness of the proposed method, we conduct comprehensive experiments over four benchmark datasets: Amazon-Beauty (Beauty for short), Amazon-Clothing (Clothing for short), LastFM, and MovieLens 1M (ML1M for short). 
The first two datasets\footnote{\url{https://nijianmo.github.io/amazon/}} are obtained from the \emph{amazon.com} e-commerce platform, encompassing a broad spectrum of user interactions with Beauty and Clothing products.
The LastFM\footnote{\url{https://grouplens.org/datasets/hetrec-2011/}} dataset provides music artist listening records from users at the Last.fm online music system. 
The ML1M\footnote{\url{https://grouplens.org/datasets/movielens/1m/}} dataset offers a collection of movie ratings made by MovieLens users.
Table~\ref{tab:dataset} shows the statistics of these four datasets, where the maximum item sequence length is set to 100 to accommodate the input length of the LLM backbone T5 (512 tokens).
Moreover, our training-validation-testing combination follows the leave-one-out policy~\cite{geng2022recommendation}, i.e., using all but the last observation in users' interaction history as the training set.
Last but not least, we randomly shuffle users' interaction history to provide a non-chronological item list to align with the settings of collaborative filtering methods.

\begin{table}[htbp]
\centering
\caption{Basic statistics of benchmark datasets.}
    \vskip -0.05in
\label{tab:dataset}
\scalebox{1}
{
\begin{threeparttable}
\begin{tabular}{c|c|c|c|c}
\toprule
\multirow{2}{*}{\textbf{Datasets}} & \multicolumn{4}{c}{\textbf{User-Item Interaction}}           \\
                                   & \textbf{\#Users} & \textbf{\#Items} & \textbf{\#Interactions} & \textbf{Density (\%)} \\ 
                                   \midrule
\textbf{LastFM} & 1,090  & 3,646 & 37,080 & 0.9330 \\
\textbf{ML1M} & 6,040 & 3,416 & 447,294 & 2.1679 \\
\textbf{Beauty}  & 22,363 & 12,101  & 197,861 & 0.0731 \\
\textbf{Clothing} & 23,033 & 39,387 & 278,641 & 0.0307 \\
\bottomrule
\end{tabular}
    \begin{tablenotes}
    \footnotesize
    \item \# represents the number of users, items, and interactions. 
    \end{tablenotes}
    \end{threeparttable}
}
\vskip -0.1in
\end{table}

\subsubsection{\textbf{Baselines}}
Here, we compare our approach with four representative collaborative filtering methods (i.e., MF, NeuCF, LightGCN,  GTN, and LTGNN), three widely-used sequential recommendation methods (i.e., SASRec, BERT4Rec, and S$^3$Rec), and five state-of-the-art LLM-based recommendation methods(i.e., P5, CID, POD, TIGER, and CoLLM).

\begin{itemize}[leftmargin=*] 

\item \textbf{Collaborative Filtering (CF) methods:} \textbf{MF}~\cite{Rendle2009BPRBP} is the most classic CF method, while \textbf{NeuCF}~\cite{he2017neural} is the very first DNN-based model. \textbf{LightGCN}~\cite{he2020lightgcn} and \textbf{GTN}~\cite{fan2022graph} are representative CF based on GNNs techniques. 
\textbf{LTGNN}~\cite{zhang2024ltgnn} is the most advanced GNN-based collaborative filtering method.

\item \textbf{Sequential Recommendations}: \textbf{SASRec}~\cite{kang2018sasrec} is an attention-based sequential recommendation model. \textbf{BERT4Rec}~\cite{sun2019bert4rec} is a bidirectional Transformer-based recommender trained with the BERT-style cloze task. \textbf{S$^3$Rec}~\cite{zhou2020s3} is a representative sequential recommendation model trained by self-supervised learning.
{\textbf{CoSeRec}~\cite{liu2021contrastive} incorporates contrastive Self-Supervised Learning to sequential recommendation.}

\item \textbf{P5}~\cite{geng2022recommendation} is a pioneering work on LLM-based RecSys, which describes recommendation tasks in a text-to-text format and employs LLMs to capture deeper semantics for personalization and recommendation. In our experiments, we deploy two indexing methods, i.e., \emph{random indexing (RID)} and \emph{sequential indexing (SID)}, on the P5 model.
Among these, P5-SID is chosen as a baseline for our efficiency evaluation, generalizability evaluation, and ablation study in our experiments.

\item \textbf{CID}~\cite{hua2023index} is an indexing approach that considers the co-occurrence matrix of items to design numeric IDs so that items co-occur in user-item interactions will have similar numeric IDs. To be consistent, we employ the P5 model as its LLM backbone.

\item \textbf{POD}~\cite{li2023prompt} encodes discrete prompts into continuous embeddings to reduce the excessive input length of LLMs based on P5  architecture.

\item \textbf{TIGER}~\cite{rajput2023recommender} condenses extensive textual data into a few semantic IDs through residual vector quantization and trains a Transformer-based model using the sequences of semantic IDs for sequential recommendation tasks. A variant is also included in our evaluation, denoted by \textbf{TIGER-G}. It is trained on collaborative IDs instead of semantic IDs. The collaborative IDs are quantized from the collaborative embeddings learned by a GNN method, specifically LightGCN.

\item \textbf{CoLLM}~\cite{zhang2023collm} employs GNNs to provide continuous embeddings representing items and users for LLM-based recommendations.
In our experiments, CoLLM's binary classification output (i.e., whether or not a user likes an item) is reformatted to generate item IDs consistent with the output setting of top-$K$ recommendations.  
\end{itemize}

\subsubsection{\textbf{Evaluation Metrics}}
In order to evaluate the quality of recommendation results, we adopt two widely used evaluation metrics: top-$K$ Hit Ratio (HR@$K$) and top-$K$ Normalized Discounted Cumulative Gain (NDCG@$K$)~\cite{wang2019neural,zhang2024ltgnn}, in which higher values indicate better performance for recommendations.
And the average metrics for all users in the test set are reported.
In addition, we set the values of $K$ as 10, 20, and 30, among which 20 is the default value for ablation experiments.
 
\subsubsection{\textbf{Hyper-parameter Settings}}
Our proposed model is implemented based on Hugging Face and PyTorch.
The codebook number $K$ (i.e., the subencoder), the token number $L$ at each subcodebook, and the ratio $\rho$ of our masking operation are search in the ranges of $\{1, 2, 3, 4, 5\}$, $\{128, 256, 512, 1024\}$, and \{0 to 1\} in 0.1 increments, respectively. 
Moreover, during the fine-tuning process of our proposal LLM-based recommendation framework, the ratio of negative sampling $\lambda$ presented in Eq.\eqref{eq:llmloss} is fixed at 1:1, in which we randomly select an un-interacted item from the whole item base as the negative sample for each positive sample. 
The margin $\gamma$ in Eq.\eqref{eq:llmloss} is set to 0 to 0.2.  
We optimize the \ourvqname{}s and the LLM backbone with AdamW~\cite{aloshchilov2019adamw} in a mini-batch manner, with a batch size of 128 and a maximum of 100 training epochs.
Note that the high-order collaborative representations for users and items are obtained from a representative collaborative filtering method, namely LightGCN~\cite{he2020lightgcn}.
For prompting, we design 11 templates for \ourname{}: 10 of them are set to be \emph{seen} prompts, and the remaining one is evaluated as an \emph{unseen} prompt.
For a fair comparison, we employ a widely-used lightweight LLM, i.e., \emph{T5-small}~\cite{raffel2020t5}, for \ourname{} and all LLM-based baselines. 
The other default hyper-parameters for baseline methods are set as suggested by the corresponding papers.
{All the experiments are conducted on a single NVIDIA A800 GPU (80 GB).}

\begin{table*}[htbp]
  \centering
  \caption{Performance comparison of recommendation algorithms on the LastFM and ML1M datasets.}
  \vskip -0.1in
\scalebox{0.95}
{
\begin{threeparttable}
    \begin{tabular}{c|cccccc|cccccc}
    \toprule
    \multirow{2}[2]{*}{Model} & \multicolumn{6}{c}{\textbf{LastFM}}           & \multicolumn{6}{c}{\textbf{ML1M}} \\
               &  HR@10   &  HR@20   &  HR@30   &  NG@10  &  NG@20  &  NG@30  &  HR@10   &  HR@20   &  HR@30   &  NG@10  &  NG@20  &  NG@30 \\
               \midrule
        BERT4Rec  & 0.0319 & 0.0461 & 0.0640 & 0.0128 & 0.0234 & 0.0244 & 0.0779 & 0.1255 & 0.1736 & 0.0353 & 0.0486 & 0.0595 \\
        SASRec  & 0.0345 & 0.0484 & 0.0658 & 0.0142 & 0.0236 & 0.0248 & 0.0785 & 0.1293 & 0.1739 & 0.0367 & 0.052 & 0.0622 \\  
        S$^3$Rec  & 0.0385 & 0.0490 & 0.0689 & 0.0177 & 0.0266 & 0.0266 & 0.0867 & 0.1270 & 0.1811 & 0.0361 & 0.0501 & 0.0601 \\
        CoSeRec & 0.0388  & 0.0504  & 0.0720  & 0.0180  & 0.0268  & 0.0278  & 0.0795  & 0.1316  & 0.1804  & 0.0375  & 0.0529  & 0.0652  \\
        \midrule
        MF     & 0.0239 & 0.0450 & 0.0569 & 0.0114 & 0.0166 & 0.0192 & 0.078 & 0.1272 & 0.1733 & 0.0357 & 0.0503 & 0.0591 \\
        NCF    & 0.0321 & 0.0462 & 0.0643 & 0.0141 & 0.0252 & 0.0254 & 0.0786 & 0.1273 & 0.1738 & 0.0363 & 0.0504 & 0.0601 \\
        LightGCN  & 0.0385 & 0.0661 & 0.0982 & 0.0199 & 0.0269 & 0.0336 & 0.0877 & 0.1288 & 0.1813 & 0.0374 & 0.0509 & 0.0604 \\
        GTN    & 0.0394 & 0.0688 & 0.0963 & 0.0199 & 0.0273 & 0.0331 & 0.0883 & 0.1307 & 0.1826 & 0.0378 & 0.0512 & 0.0677 \\
        LTGNN  & 0.0471 & 0.076 & 0.0925 & 0.0234 & 0.0318 & 0.0354 & 0.0915 & 0.1387 & 0.1817 & 0.0419 & 0.0570 & 0.0659 \\
        \midrule
        P5-RID  & 0.0312 & 0.0523 & 0.0706 & 0.0144 & 0.0199 & 0.0238 & 0.0867 & 0.1248 & 0.1811 & 0.0381 & 0.0486 & 0.0662 \\
        P5-SID  & 0.0375 & 0.0536 & 0.0851 & 0.0224 & 0.0255 & 0.0261 & 0.0892 & 0.1380 & 0.1784 & 0.0422 & 0.0550 & 0.0641 \\
         CID  & 0.0381 & 0.0552 & 0.0870 & 0.0229 & 0.0260 & 0.0277 & 0.0901 & 0.1294 & 0.1863 & 0.0379 & 0.0525 & 0.0706 \\
        POD    & 0.0367 & 0.0572 & 0.0747 & 0.0184 & 0.0220 & 0.0273 & 0.0886 & 0.1277 & 0.1846 & 0.0373 & 0.0487 & 0.0668 \\
        TIGER & 0.0467 & 0.0749 & 0.0984 & 0.0226 & 0.0306 & 0.0348 & 0.0901 & 0.1382 & 0.1803 & 0.0427 & 0.0562 & 0.0653 \\
        TIGER-G & 0.0470 & 0.0767 & 0.0997 & 0.0229 & 0.031 & 0.0355 & 0.0905 & 0.1409 & 0.1824 & 0.0423 & 0.0565 & 0.0651 \\
        CoLLM  & 0.0483 & 0.0786 & 0.1017 & 0.0234 & 0.0319 & 0.0366 & 0.0923 & 0.1499 & 0.1998 & 0.0456 & 0.0620 & 0.0719 \\
        \midrule
        * (User ID Only)  & 0.0505 & 0.0881 &  \underline{0.1128}  &  \underline{0.0251}  &  \underline{0.0345}  & 0.0397 & 0.0964 & 0.1546 & 0.2043 & 0.0493 & 0.0640 & 0.0745 \\
        * (Unseen Prompt)  &  \underline{0.0514}  &  \underline{0.0917}  &  \textbf{0.1294}  &  \textbf{0.0252}  & 0.0343 &  \textbf{0.0422}  &  \textbf{0.1012}  &  \underline{0.1672}  &  \underline{0.2144}  &  \textbf{0.0532}  &  \textbf{0.0698}  & \textbf{0.0798} \\
        \ourname{}  &  \textbf{0.0532}  &  \textbf{0.0936}  & 0.1248 & 0.0247 &  \textbf{0.0348}  &  \underline{0.0415}  &  \underline{0.1008}  &  \textbf{0.1677}   &  \textbf{0.2149}  &  \underline{0.0528}  &  \underline{0.0697}   & \underline{0.0797} \\

    \bottomrule
    \end{tabular}%
    \begin{tablenotes}
    \footnotesize
    \item * are the variants of \textbf{\ourname{}}, namely the cases of using user ID tokens only for model inputs without considering item interaction history and using the unseen prompt during evaluation. 
    \end{tablenotes}
    \end{threeparttable}
    }
  \label{tab:comparison_LastFM&ML1M}%
\end{table*}%

\begin{table*}[htbp]
  \centering
  \caption{Performance comparison of recommendation algorithms on the Beauty and Clothing datasets.}
  \vskip -0.1in
  \scalebox{0.95}
{
\begin{threeparttable}
    \begin{tabular}{c|cccccc|cccccc}
    \toprule
    \multirow{2}[2]{*}{Model} & \multicolumn{6}{c}{\textbf{Beauty}}           & \multicolumn{6}{c}{\textbf{Clothing}} \\
  &  HR@10   &  HR@20   &  HR@30   &  NG@10  &  NG@20  &  NG@30  &  HR@10   &  HR@20   &  HR@30   &  NG@10  &  NG@20  & NG@30 \\
  \midrule
        BERT4Rec  & 0.0329 & 0.0464 & 0.0637 & 0.0162 & 0.0205 & 0.0255 & 0.0135 & 0.0217 & 0.0248 & 0.0061 & 0.0074 & 0.0079 \\
        SASRec  & 0.0338 & 0.0472 & 0.0637 & 0.0170 & 0.0213 & 0.0260 & 0.0136 & 0.0221 & 0.0256 & 0.0063 & 0.0076 & 0.0081 \\ 
        S$^3$Rec  & 0.0351 & 0.0471 & 0.0664 & 0.0169 & 0.0237 & 0.0278 & 0.0140 & 0.0213 & 0.0256 & 0.0069 & 0.0081 & 0.0086 \\
        CoSeRec & 0.0362  & 0.0476  & 0.0680  & 0.0176  & 0.0248  & 0.0280  & 0.0139  & 0.0211  & 0.0251  & 0.0068  & 0.0080  & 0.0085  \\
        \midrule
        MF     & 0.0127 & 0.0195 & 0.0245 & 0.0063 & 0.0081 & 0.0091 & 0.0116 & 0.0175 & 0.0234 & 0.0074 & 0.0088 & 0.0101 \\
        NCF    & 0.0315 & 0.0462 & 0.0623 & 0.0160 & 0.0196 & 0.0237 & 0.0119 & 0.0178 & 0.024 & 0.0072 & 0.0090 & 0.0103 \\
        LightGCN  & 0.0344 & 0.0498 & 0.0630 & 0.0194 & 0.0233 & 0.0261 & 0.0157 & 0.0226 & 0.0279 & 0.0085 & 0.0103 & 0.0114 \\
        GTN    & 0.0345 & 0.0502 & 0.0635 & 0.0198 & 0.0241 & 0.0268 & 0.0158 & 0.0226 & 0.0282 & 0.0084 & 0.0103 & 0.0111 \\
        LTGNN  & 0.0385 & 0.0564 & 0.0719 & 0.0207 & 0.0252 & 0.0285 & 0.0155 & 0.0218 & 0.0272 & 0.0082 & 0.0110 & 0.0116 \\
        \midrule
        P5-RID  & 0.0330 & 0.0511 & 0.0651 & 0.0146 & 0.0200  & 0.0144 & 0.0148 & 0.0225 & 0.0263 & 0.0071 & 0.0086 & 0.0095 \\
        P5-SID  & 0.0340 & 0.0516 & 0.0672 & 0.0154 & 0.0231 & 0.0176 & 0.0143 & 0.0222 & 0.0258 & 0.0070 & 0.0086 & 0.0091 \\
        CID  & 0.0341 & 0.0516 & 0.0673 & 0.0165 & 0.0236 & 0.0177 & 0.0146 & 0.0226 & 0.0276 & 0.0070 & 0.0087 & 0.0092 \\
        POD    & 0.0339 & 0.0498 & 0.0639 & 0.0185 & 0.0222 & 0.0221 & 0.0147 & 0.0225 & 0.0261 & 0.0074 & 0.0087 & 0.0091 \\
        TIGER & 0.0372  & 0.0574  & 0.0747  & 0.0193  & 0.0248  & 0.0287  & 0.0147  & 0.0225  & 0.0266  & 0.0072  & 0.0087  & 0.0093  \\
        TIGER-G & 0.0382  & 0.0586  & 0.0753  & 0.0195  & 0.0251  & 0.0292  & 0.0147  & 0.0227  & 0.0265  & 0.0073  & 0.0088  & 0.0093  \\
        CoLLM  & 0.0391 & 0.0606 & 0.0772 & 0.0200  & 0.0259 & 0.0303 & 0.0150 & 0.0218 & 0.0274 & 0.0079 & 0.0091 & 0.0117 \\
        \midrule
        * (User ID Only)  & 0.0396 & 0.0599 & 0.0763 & 0.0214 & 0.0265 & 0.0300  & 0.0160 & 0.0228 & 0.0282 & 0.0092 & 0.0109 & 0.0119 \\
        * (Unseen Prompt)  &  \underline{0.0402}  &  \textbf{0.0622}  &  \textbf{0.0791}  &  \underline{0.0215}  &  \underline{0.0270}  &  \textbf{0.0306}  &  \underline{0.0164}  &  \underline{0.0233}  &  \underline{0.0286}  &  \underline{0.0096}  &  \underline{0.0111}  & \underline{0.0124} \\
        \ourname{}  &  \textbf{0.0407}   &  \underline{0.0615}   &  \underline{0.0782}   &  \textbf{0.0222}   &  \textbf{0.0276}   &  \underline{0.0303}   &  \textbf{0.0171}   &  \textbf{0.0240}   &  \textbf{0.0291}   &  \textbf{0.0108}   &  \textbf{0.0112}  & \textbf{0.0130} \\
    \bottomrule
    \end{tabular}%
    \begin{tablenotes}
    \footnotesize
    \item * are the variants of \textbf{\ourname{}}, namely the cases of using user ID tokens only for model inputs without considering item interaction history and using the unseen prompt during evaluation. 
    \end{tablenotes}
    \end{threeparttable}
}
\label{tab:comparison_Beauty&Clothing}%
\vskip -0.2in
\end{table*}%

\subsection{Performance Comparison of Recommender Systems}
We first compare the recommendation performance between \ourname{} and all baselines over four benchmark datasets. 
Table~\ref{tab:comparison_LastFM&ML1M} and Table~\ref{tab:comparison_Beauty&Clothing} present the overall performance comparison on the four datasets, where * denotes the proposed \ourname{}, and the best and second best results are marked by \textbf{Bold} and \underline{underlined}.
Notably, the proposed model using unseen prompts will be indicated by the suffix (Unseen Prompt).
Besides, the suffix of (User ID Only) denotes the case of using \textbf{user ID tokens} only for LLM-empowered recommendation without considering users' interaction history towards items. 
We make the following observations:

\begin{itemize}[leftmargin=*] 
\item   \vspace{-0.05in}
Our proposed \ourname{} achieves the best performance and consistently outperforms all the baselines across all datasets regarding metrics with either unseen personalized prompts.
On average, \ourname{} significantly exceeds the strongest baselines by 19.08\% on HR@20 and 9.09\% on NCDG@20 in the LastFM dataset.
Such improvement demonstrates the effectiveness of our proposed method and the great potential of exploring collaborative indexing (i.e., tokenization) in LLM-based RecSys.

\item Even when using only user ID tokens, \ourname{} surpasses most baselines in terms of accuracy, indicating its superior performance in collaborative recommendations.
Such a result implies that \ourname{} is capable of effectively modeling user even in the absence of their interaction history. 
Rather than spending a large number of tokens describing users' interacted items, this allows \ourname{} to generate recommendations using a concise input, thus circumventing the input length restrictions imposed by LLMs and saving considerable computation resources.

\item As an earlier non-trivial indexing method to capture hidden knowledge in co-occurrence frequency, Collaborative Indexing (CID) outperforms Random Indexing (RID) and Sequential Indexing (SID) given the same setting of P5.
These observations suggest the potential of integrating collaborative knowledge for item\&user tokenization/indexing.
However, the P5 variants and POD are inferior to the existing GNN-based collaborative filtering (i.e., LightGCN, GTN, and LTGNN), implying their inability to capture collaborative information using LLMs.
In constrast, CoLLM achieves superior results, benefiting from its incorporation of collaborative embeddings learned from these GNN methods.

\item Thanks to its use of residual vector quantization and semantic IDs for item indexing, TIGER outperforms typical LLM-based recommendation methods, such as P5-RID and P5-SID. Moreover, the variant TIGER-G demonstrates improved performance in many cases compared to TIGER, further highlighting the benefits of incorporating collaborative knowledge in LLM-based recommendation tasks.

\item {The GNN-based collaborative filtering methods perform relatively better than the traditional CF methods (i.e., MF and NCF) and representative sequential recommendation methods (i.e., BERT4Rec, SASRec, S$^3$Rec, and CoSeRec).} The results demonstrate the effectiveness of GNNs in capturing collaborative signals via high-order connectivity.
\end{itemize}

\subsection{Generalizability Evaluation}
\label{subsec:Generalizability}
In most e-commerce and social media platforms, a significant number of new users and items are added daily to recommender systems.
As a result, well-established systems are required to frequently conduct updates and enhancements to accommodate and generalize to the preferences of new users and the characteristics of new items, thereby providing personalized recommendations that cater to the evolving dynamics of their user base. 
Since newly added users and items lack their interactions in fine-tuning LLM4Rec, most existing LLM-based RecSys fail to retrieve suitable items as potential candidates, thus requiring extensive retraining. 
In contrast, our proposed \ourname{} can easily generalize effectively even in cases where the user is not present in the training and fine-tuning corpus.

For this analysis, we consider the Beauty and LastFM datasets and {exclude the 5\% of users (ref as \emph{unseen users}) with the least interaction history from the training data split to simulate newly added users.
Consequently, all items interacted with by these users are also excluded from the training set.}
The performance of \ourname{} is compared to other LLM-based recommendation models, i.e., P5, POD, CID, TIGER, and CoLLM.
To ensure the absence of data leakage concerning unseen users, only the training split is utilized for establishing these models, while it is allowed to update the vector database to provide collaborative representations of unseen users and items (as shown in Figure~\ref{fig:inference}), which consumes far less computational resources than updating the LLM backbone.
From the recommendation results presented in Table~\ref{tab:generalizability}, several key observations can be made:
\begin{itemize}[leftmargin=*] 
    \item Existing LLM-based recommendation methods face challenges in generalizability, as evidenced by a significant drop of over 40\% in HR@20 and NDCG@20 for P5 and POD when recommending items to unseen users.
    
    \item The inclusion of collaborative knowledge in CID and CoLLM leads to a relatively improved performance in model generalization, as indicated by a reduced performance degradation in P5 and POD.
    However, these methods still experience more than a 20\% drop, indicating that they overlook the importance of stable ID tokenization for LLM-based recommendations.
    
    \item In comparison, \ourname{} outperforms the aforementioned methods for not only the training users but also the unseen users in both datasets.
    For instance, in the Amazon-Beauty dataset, the performance of \ourname{} decreases by only 7\% on average, demonstrating the strong generalization capability of \ourname{} for newly added users.
    Such a superiority can be attributed to our \ourvqname{} for robust ID tokenization and the generative retrieval paradigm for flexible recommendation generation.

    \item Additionally, TIGER exhibits strong generalization capability for unseen cases, with only a 6.10\% average decrease observed on HR@20. This performance can be attributed to the use of semantic IDs, which incorporate item-side textual information as additional knowledge—a feature not present in other compared methods. This approach holds potential for addressing cold-start scenarios by incorporating item-side information effectively.
\end{itemize}

\begin{table}[htbp]
  \centering
  \caption{Performance comparison on seen and unseen users for generalizability evaluation.}
    \vskip -0.05in
    \begin{tabular}{c|c|cc|cc}
    \toprule
    \multirow{2}[2]{*}{Dataset} & \multirow{2}[2]{*}{Model} & \multicolumn{2}{c}{Seen} & \multicolumn{2}{c}{Unseen} \\
          &       & HR@20 & NG@20 & HR@20 & NG@20 \\
    \midrule
    \multirow{5}[2]{*}{LastFM} 
          & P5 & 0.0704  & 0.0320  & 0.0399  & 0.0137  \\
    & POD   & 0.0709  & 0.0323  & 0.0401  & 0.0138  \\
    & CID   & 0.0697  & 0.0314  & 0.0452  & 0.0196  \\
    & TIGER & 0.0752 & 0.0309 & \underline{0.0695} & \underline{0.0252} \\
    & CoLLM & \underline{0.0812}  & \underline{0.0336}  & 0.0574  & 0.0235  \\
          & \ourname{} & \textbf{0.0973}  & \textbf{0.0353}  & \textbf{0.0773}  & \textbf{0.0268}  \\
    \midrule
    \multirow{5}[2]{*}{Beauty} 
         & P5 & 0.0511  & 0.0236  & 0.0274  & 0.0130  \\
   &  POD   & 0.0507  & 0.0225  & 0.0269  & 0.0123  \\
   &  CID   & 0.0523  & 0.0240  & 0.0334  & 0.0146  \\
   & TIGER & 0.0575 & 0.0248 & \underline{0.0548} & \underline{0.0233} \\
   &  CoLLM & \underline{0.0612}  & \underline{0.0261}  & 0.0477  & 0.0195  \\
          & \ourname{} & \textbf{0.0629}  & \textbf{0.0289}  & \textbf{0.0591}  & \textbf{0.0266}  \\ 
    \bottomrule
    \end{tabular}%
  \label{tab:generalizability}%
  \vskip -0.2in
\end{table}%

\subsection{Efficiency Evaluation}
\label{subsec:Efficiency}
In this subsection, we analyze the inference efficiency of \ourname{} compared with the representative LLM-based recommendation methods (i.e., P5, CID, POD, and {TIGER}), which generate the tokens (e.g., IDs, titles, and descriptions) for Top-$K$ items by using typical auto-regressive decoding and beam search. 
Our proposed method discards this text-decoding generation solution and introduces a generative retrieval paradigm to perform collaborative recommendations.
{This evaluation is conducted under the same default settings as those used in the overall performance comparisons.
The top-20 retrieval results, as presented in Table~\ref{tab:efficient}, demonstrate that TIGER outperforms its text-based counterparts (i.e., P5, CID, and POD) in terms of efficiency, as it utilizes compressed semantic IDs, which require fewer tokens than item titles or descriptions.
More importantly, \ourname{} can achieve superior inference efficiency with a significant improvement of approximately 1259.81\% compared to the baselines.
This can be attributed to our generative retrieval paradigm, bypassing the most time-consuming auto-regressive decoding and beam search processes of LLMs~\cite{wang2024rethinking}.}

\begin{table}[htbp]
  \centering
  \caption{ {Comparison of the average inference time per user (in milliseconds) for Top-20 recommendations.} }
    \vskip -0.05in
  \begin{threeparttable}
    \begin{tabular}{ccccc}
    \toprule
    Inference Time & LastFM & ML1M  & Beauty & Clothing \\
    \midrule
     P5 & 96.04 & 99.75 & 86.39 & 93.38 \\
    POD & 96.30 & 101.42 & 87.69 & 94.48 \\
    CID & 94.96 & 99.42 & 84.87 & 92.02 \\
    {TIGER} & 82.57 & 85.98 & 76.11 & 80.68\\
    \ourname{} & 6.92  & 8.43  & 5.76  & 6.00 \\
    \midrule
    Acceleration* & 1236.24\% & 1046.41\% & 1354.25\% & 1402.33\% \\
    \bottomrule
    \end{tabular}%
        \begin{tablenotes}
    \footnotesize
    \item * The average improvement compared to the baselines.
    \end{tablenotes}
    \end{threeparttable}
  \label{tab:efficient}%
  \vskip -0.2in
\end{table}%

\subsection{Ablation Study}
In order to assess the effectiveness of the proposed key components, we conducted ablation experiments on the LastFM and Amazon-Beauty datasets, where the influence of each component was eliminated separately as follows:
\begin{itemize}[leftmargin=*]
    \item w/o High-Order Collaborative Knowledge (HOCK): 
    Use Matrix Factorization (MF) to learn the collaborative representations of users and items for ID tokenization in MQ-Tokenizers.
    Compared to advanced GNN-based methods (e.g., LightGCN), it is challenging for MF to capture high-order collaborative signals explicitly among user-item interactions.
    
    \item w/o $K$-way: Replace the $K$-way encoder and codebook with $1$-way ones while keeping the total number of codebook tokens (i.e., codewords) consistent.

    \item w/o Masking: Deactivate the masking operation in our user and item \ourvqname{}s.

    \item {s K-Means, VQ-VAE, and RQ-VAE: Substitute the proposed MQ-Tokenizers with three typical vector quantization/clustering methods, specifically K-Means, VQ-VAE~\cite{van2017neural}, and RQ-VAE~\cite{lee2022autoregressive} to validate its effectiveness.}
\end{itemize}

We can have the following observations from the ablation results in Table~\ref{tab:ablation}.
First, each component in our approach contributes to the overall performance since eliminating any of them would result in performance degradation.
Second, the masking operation and the $K$-way framework introduced in our approach not only enhance the generalizability, as demonstrated in Table~\ref{tab:generalizability}, but also result in moderate performance improvements during the ablation experiment.
{Moreover, the removal of the high-order collaborative knowledge learned by lightGCN resulted in a significant drop in performance.
This indicates the importance of incorporating such knowledge for the alignment of LLMs with personalized recommendations.
Furthermore, it demonstrates that the quality of collaborative embeddings is crucial to the effectiveness of our tokenizer and model.}
Third, the comparison with the two representative quantization methods, namely VQ-VAE and RQ-VAE, illustrates the effectiveness of our MQ-Tokenizers in encoding collaborative knowledge for LLM-based recommendations, while the inferior performance of K-Means underscores the necessity for a more robust quantization approach.

\begin{table}[htbp]
  \centering
  \caption{{Results of Ablation Studies.}}
  \vskip -0.05in
  \begin{threeparttable}
    \begin{tabular}{lcccc}
    \toprule
    \multirow{2}[2]{*}{Module} & \multicolumn{2}{c}{LastFM} & \multicolumn{2}{c}{Beauty} \\
          & HR@20  & NG@20  & HR@20  & NG@20 \\
    \midrule
    Full*  & 0.0936  & 0.0348  & 0.0615  & 0.0276  \\
    \midrule
    w/o Masking & 0.0848  & 0.0332  & 0.0573  & 0.0253  \\
    w/o $K$-way & 0.0820  & 0.0309  & 0.0592  & 0.0250  \\
    {w/o HOCK} & 0.0549  & 0.0172  & 0.0407  & 0.0149  \\
    \midrule
    s RQ-VAE & 0.0831  & 0.0314  & 0.0596  & 0.0253  \\
    {s VQ-VAE} & 0.0810  & 0.0308  & 0.0589  & 0.0247  \\
    {s K-Means} & 0.0750  & 0.0281  & 0.0567  & 0.0237  \\
    \bottomrule
    \end{tabular}%
            \begin{tablenotes}
    \footnotesize
    \item * "Full" denotes the complete version of TokenRec.
    \item {"s" denotes the substitution made to the MQ-Tokenizers.}
    \end{tablenotes}
    \end{threeparttable}
  \label{tab:ablation}%
  \vskip -0.2in
\end{table}%

\subsection{Hyper-parameter Analysis}
In \ourname{}, we introduce three critical hyper-parameters, namely the masking ratio $\rho$ in vector quantization, the number of sub-encoders/sub-codebooks $K$, and the number of tokens $L$ in each sub-codebook. 
Their value sensitivities are evaluated in this section to facilitate the future application of our proposed model.

\subsubsection{\textbf{Effect of Masking Ratio $\rho$}}
\begin{figure*}[htbp]
\centering
{\subfigure[LastFM - HR@20]
{\includegraphics[width=0.24\linewidth]{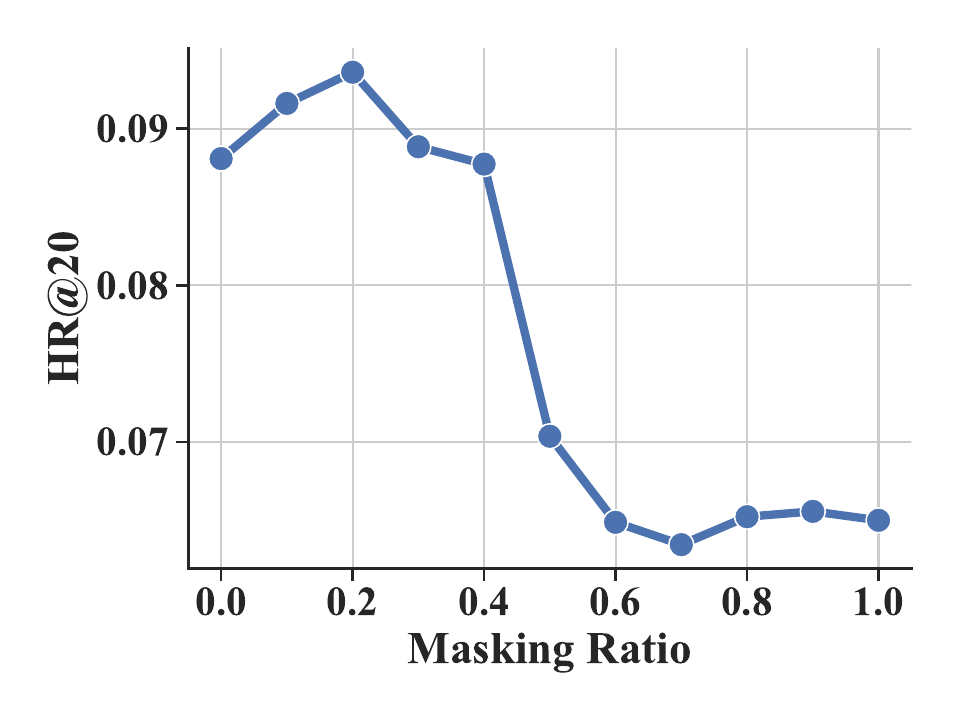}}}
{\subfigure[ML1M - HR@20]
{\includegraphics[width=0.24\linewidth]{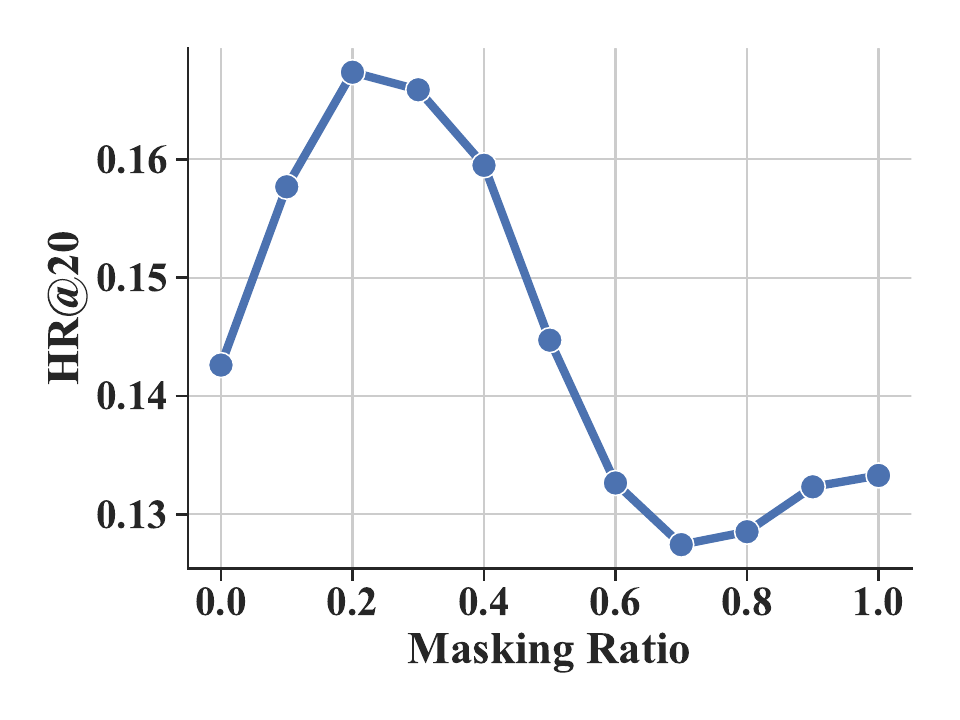}}}
{\subfigure[Amazon-Beauty - HR@20]
{\includegraphics[width=0.24\linewidth]{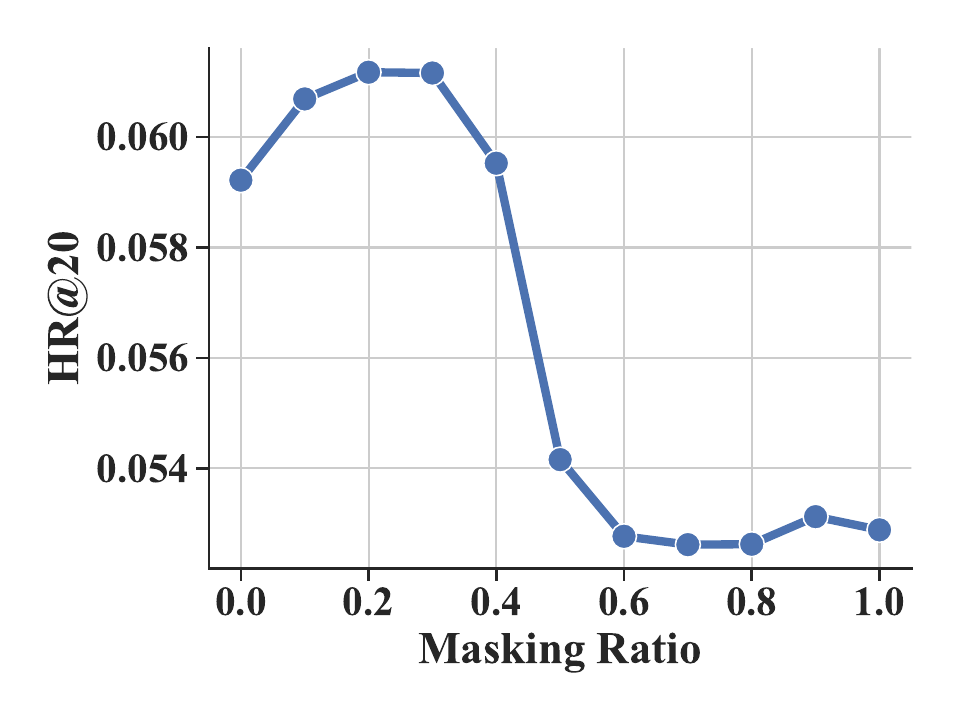}}}
{\subfigure[Amazon-Clothing - HR@20]
{\includegraphics[width=0.24\linewidth]{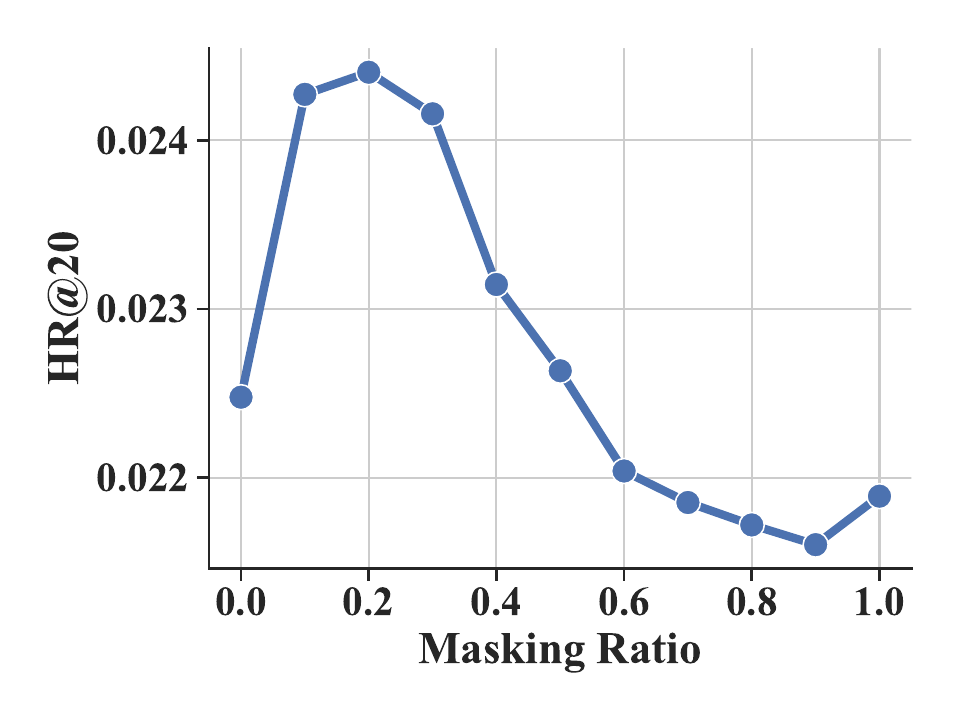}}}
{\subfigure[LastFM - NDCG@20]
{\includegraphics[width=0.24\linewidth]{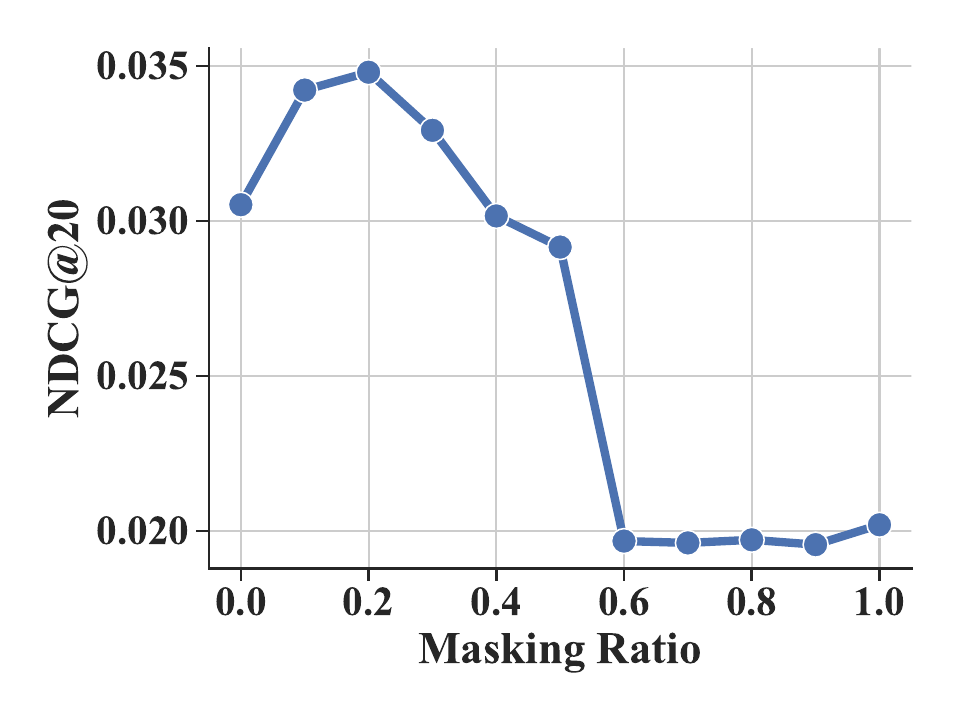}}}
{\subfigure[ML1M - NDCG@20]
{\includegraphics[width=0.24\linewidth]{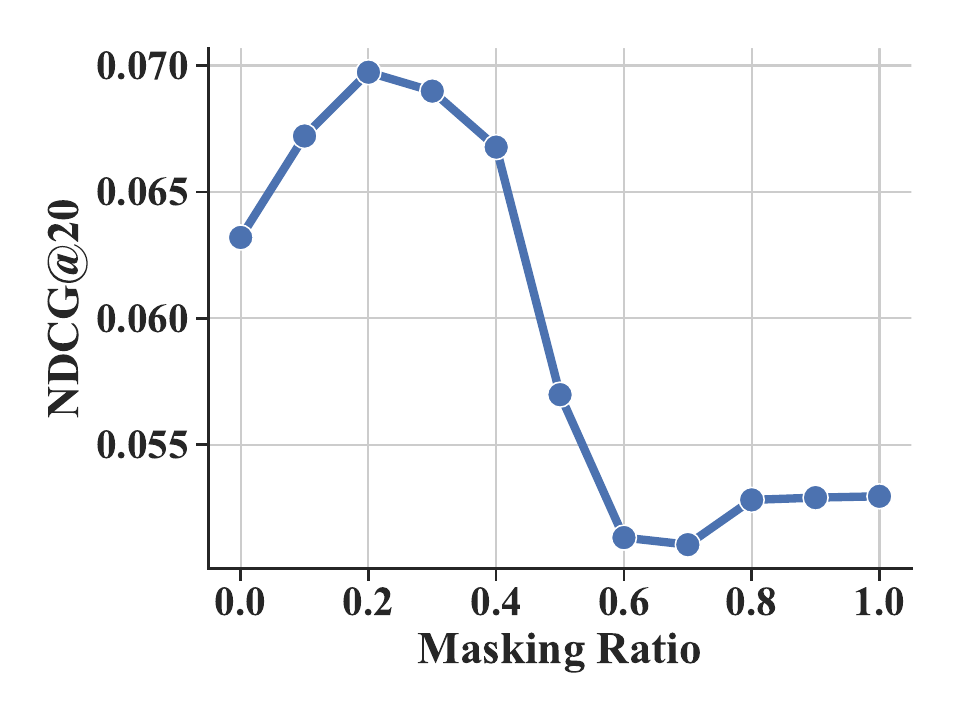}}}
{\subfigure[Amazon-Beauty - NDCG@20]
{\includegraphics[width=0.24\linewidth]{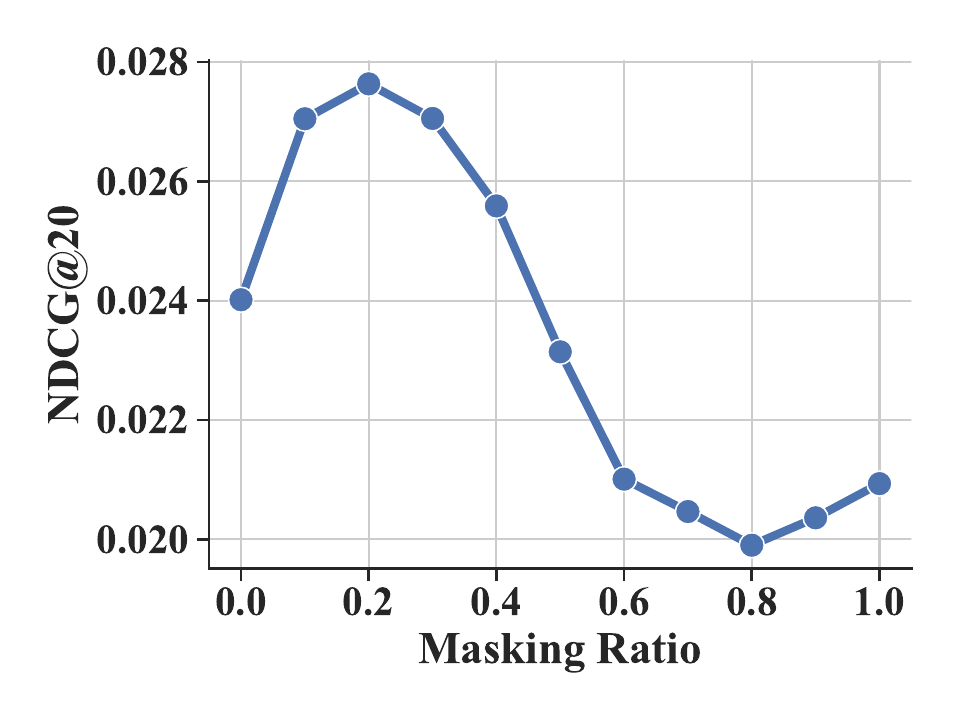}}}
{\subfigure[Amazon-Clothing - NDCG@20]
{\includegraphics[width=0.24\linewidth]{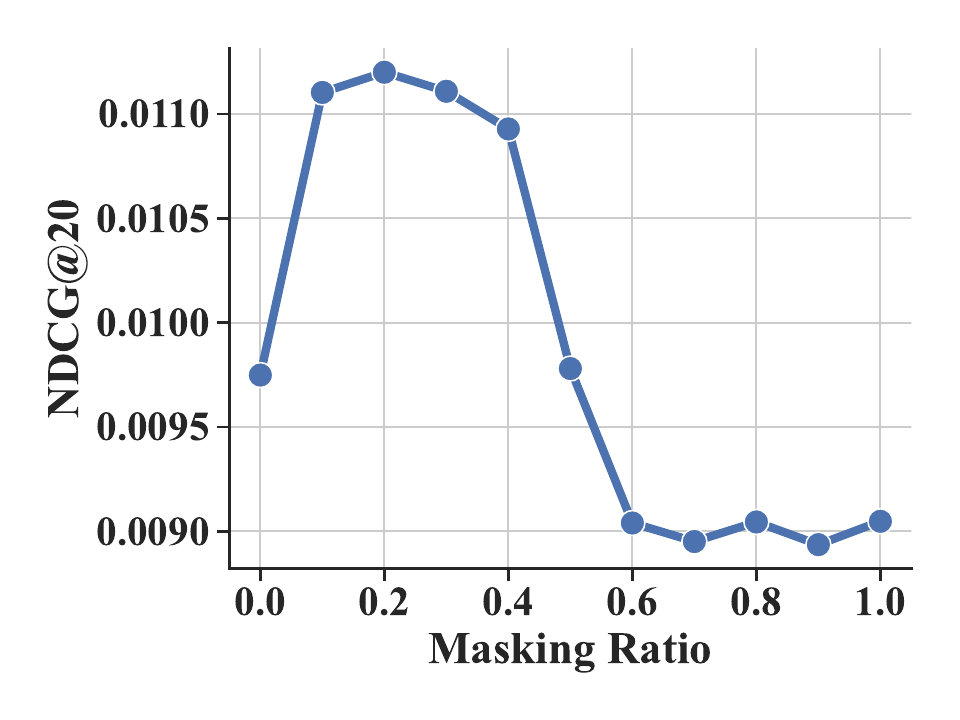}}}
\vskip -0.1in
\caption{The effect of masking ratio $\rho$ under HR@20 and NDCG@20 metrics.}\label{fig:mask}
\vskip -0.1in
\end{figure*}

We first investigate the impact of the hyper-parameter $\rho$ in \ourvqname{}, which controls the masking ratio of our vector quantization for users\& items ID tokenization. Figure~\ref{fig:mask} shows the performance change of \ourname{} w.r.t. HR@20 and NDCG@20. 
We can find that introducing a small ratio of masking brings performance improvements.
In most cases, the recommendation performance of our proposed method improves when $\rho < 0.5$, among which 0.2 can achieve the best improvement in our experiments.
The experimental results also reveal that the recommendation performance degrades when the masking ratio $\rho >= 0.5$, suggesting excessive masking should be avoided. 

\begin{figure*}[htbp]
\centering
{\subfigure[LastFM - HR@20]
{\includegraphics[width=0.24\linewidth]{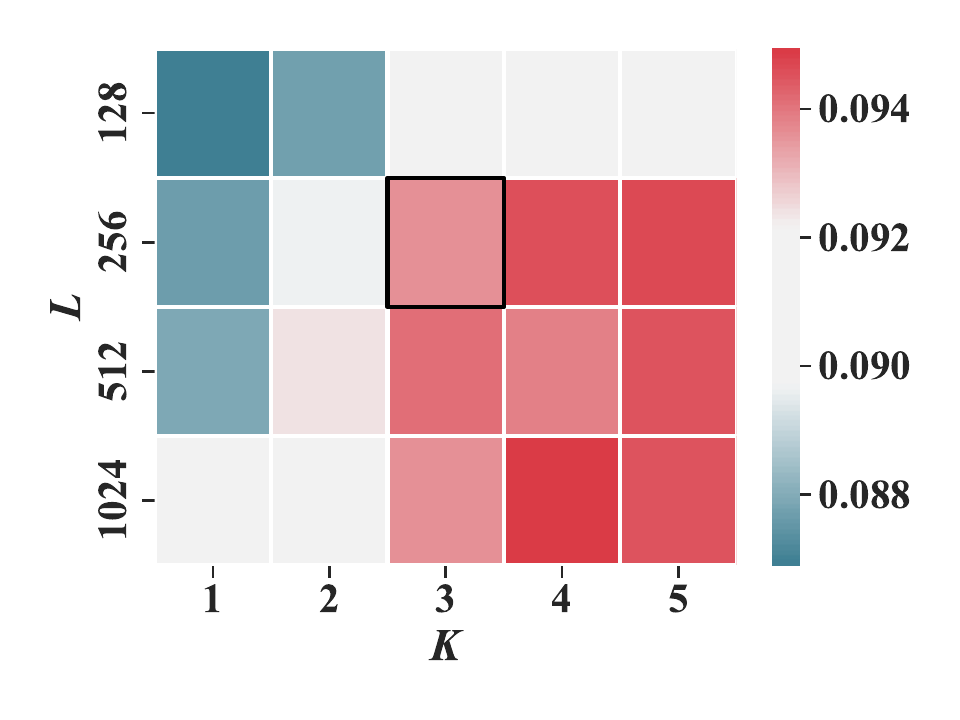}}}
{\subfigure[ML1M - HR@20]
{\includegraphics[width=0.24\linewidth]{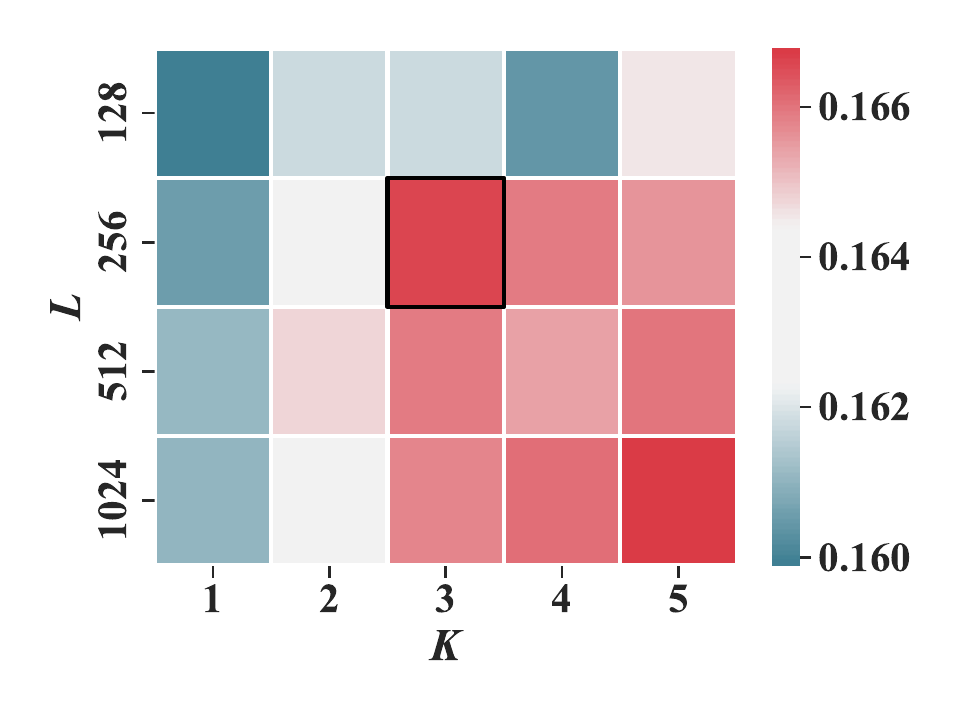}}}
{\subfigure[Amazon-Beauty - HR@20]
{\includegraphics[width=0.24\linewidth]{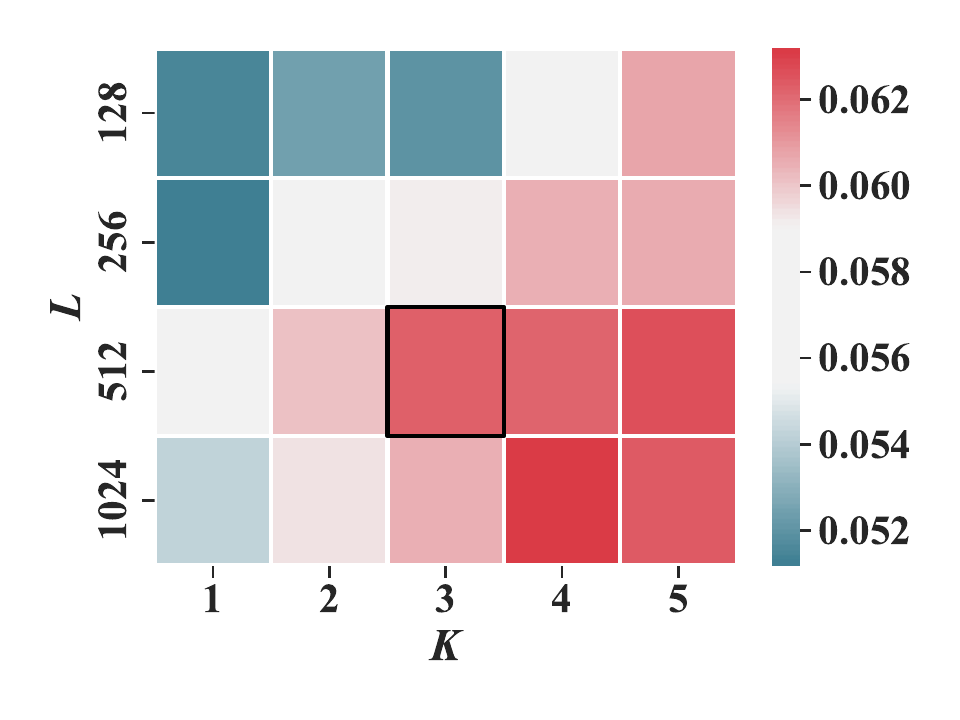}}}
{\subfigure[Amazon-Clothing - HR@20]
{\includegraphics[width=0.24\linewidth]{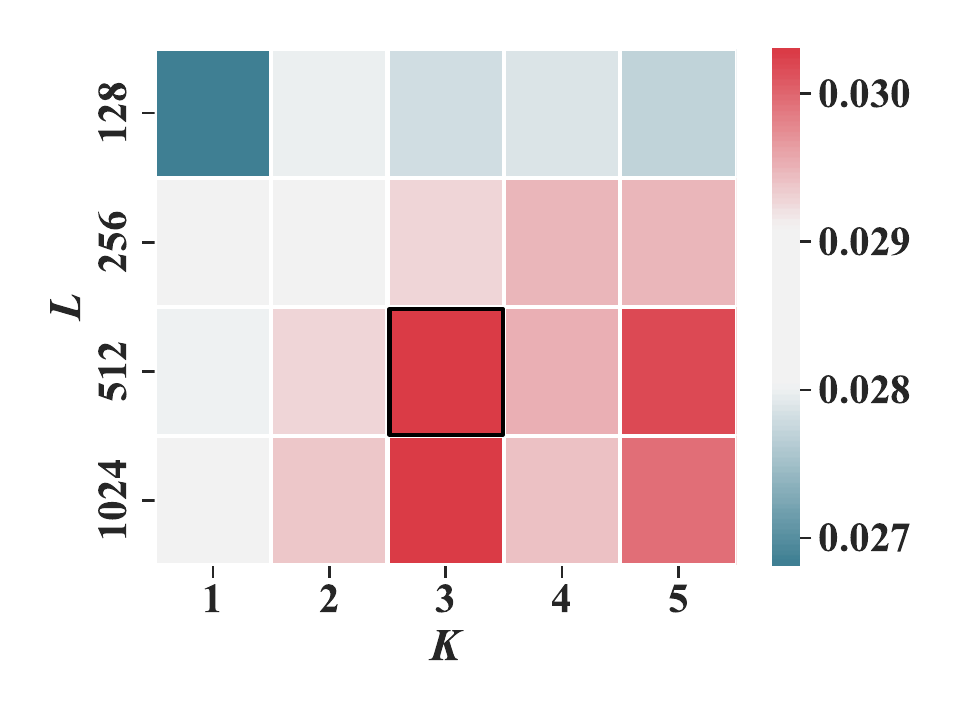}}}
{\subfigure[LastFM - NDCG@20]
{\includegraphics[width=0.24\linewidth]{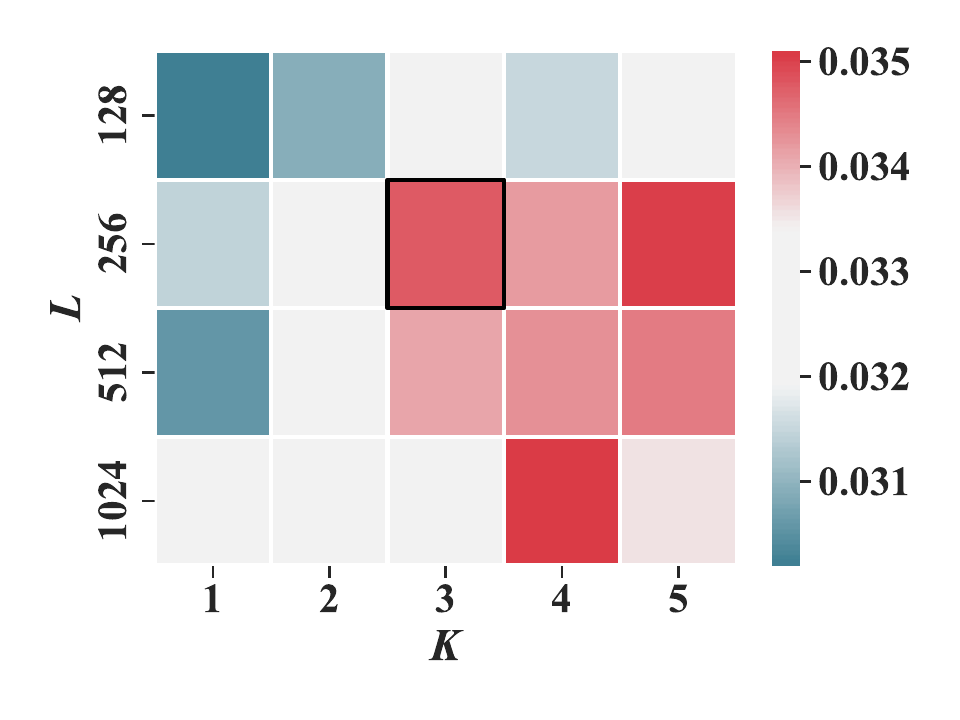}}}
{\subfigure[ML1M - NDCG@20]
{\includegraphics[width=0.24\linewidth]{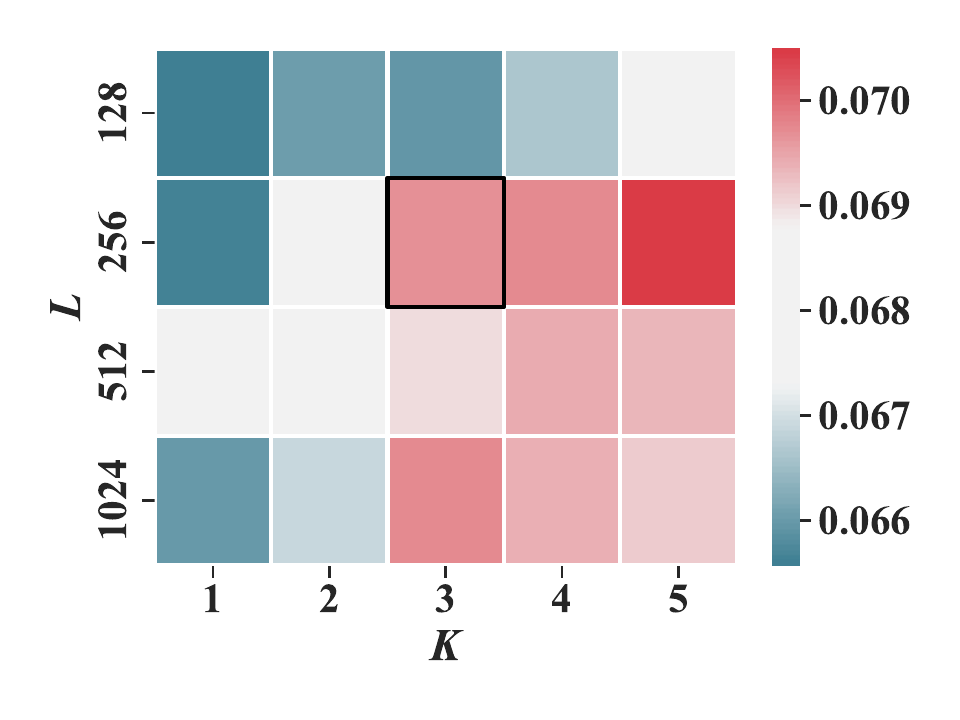}}}
{\subfigure[Amazon-Beauty - NDCG@20]
{\includegraphics[width=0.24\linewidth]{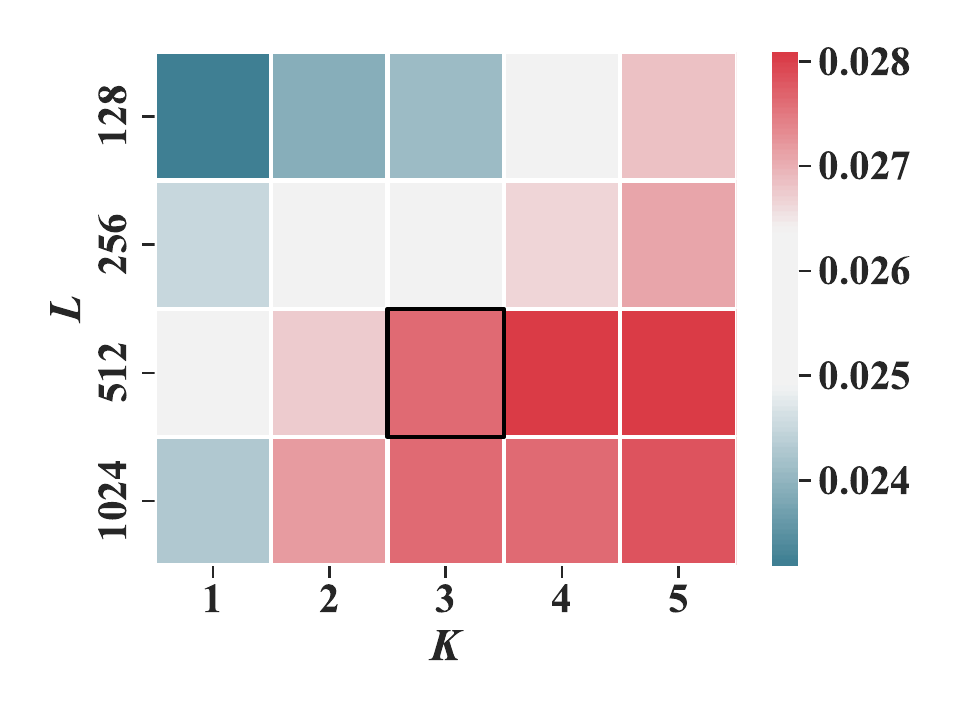}}}
{\subfigure[Amazon-Clothing - NDCG@20]
{\includegraphics[width=0.24\linewidth]{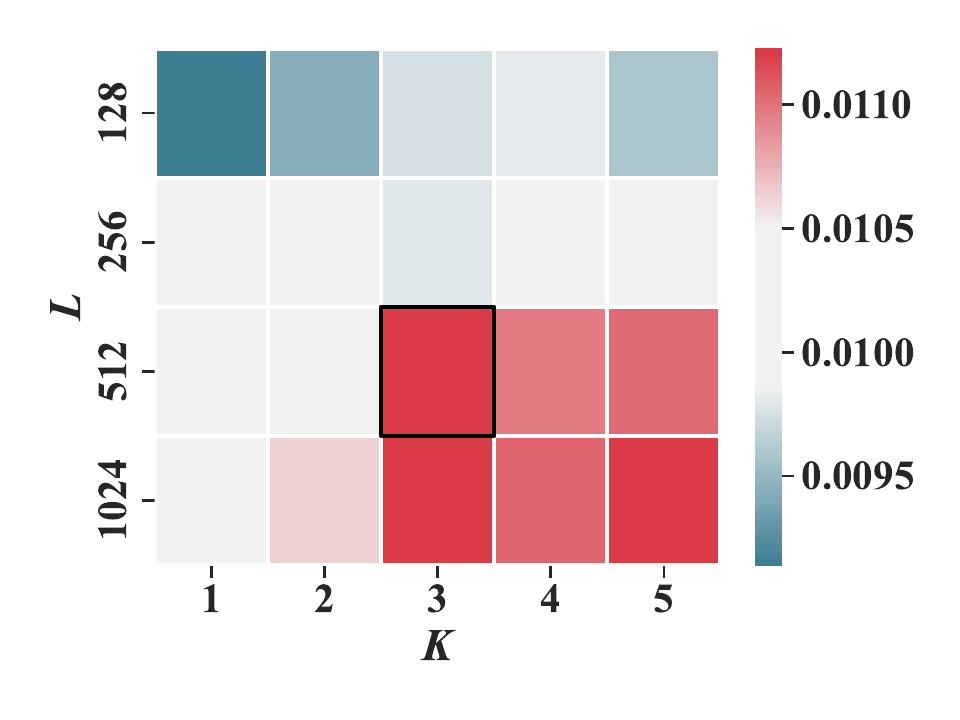}}}
\vskip -0.1in
\caption{The effect of the number of sub-codebooks $K$ and the number of tokens in each sub-codebook $J$ under HR@20 and NDCG@20 metrics.}\label{fig:hyper}
\vskip -0.1in
\end{figure*}

\subsubsection{\textbf{Effect of Codebook Settings $K$ and $L$}}
To study whether our proposed method \ourname{} can benefit from stacking more sub-codebooks and introducing more codebook tokens, we vary the numbers of the hyper-parameters $K$ and $L$ in the range of \{1, 2, 3, 4, 5\} and \{128, 256, 512, 1024\}, respectively, and report the performance on all datasets in Figure~\ref{fig:hyper}, with the optimal cases highlighted by black box.
From the figure, we can make the following observations.
\begin{itemize}[leftmargin=*]
    \item With the increase of codebook depth, a progressive performance improvement of our model for recommendations is witnessed in all datasets.
    Nevertheless, when $K>3$, the improvement of recommendation performance becomes relatively marginal.
    Thus, $K=3$ is suggested by considering the trade-off between effectiveness and efficiency.
    
    \item 
    The optimal value of $L$ varies in relation to the differing sizes of users and items. 
    Specifically, the best-balanced performance in terms of effectiveness and efficiency can be observed at a $L$ of 256 and 512 on the LastFM/ML1M (smaller sizes) and the Amazon-Beauty/Clothing (larger sizes) datasets, respectively.
    This implies that slightly more codebook tokens should be used in each sub-codebook for datasets with more users/items.
    
    \item Additionally, in the case of a single codebook, merely increasing the number of codebook tokens cannot effectively deliver performance gains in recommendations.
    This demonstrates the effectiveness of the proposed $K$-way mechanism in our \ourvqname{} for recommendations.
\end{itemize}

\section{Related Work}
\label{sec:relatedwork}
In this section, we briefly review related work, including collaborative filtering and Large Language Models (LLMs).

\subsection{Collaborative Filtering}
In order to provide personalized recommendations that accord with user preferences, collaborative filtering (CF) serves as a representative technique for modeling collaborative information in recommender systems, such as user-item interactions, to capture similar user patterns and predict future interactions~\cite{fan2021attacking,qu2024ssd4rec,fan2019deep_daso}.
As a representative example, matrix factorization (MF)~\cite{koren2009matrix,fan2018deep} vectorizes users and items into dense representations and models their interactions (i.e., user-item ID matrix) by calculating the inner products between the vectorized representations.
Subsequently, NeuCF~\cite{he2017neural} incorporates neural networks with MF to decompose the use-item interactions into two low-rank matrices representing user\&item embeddings.
Later on, DSCF~\cite{fan2019deep_dscf} takes advantage of deep language models to enhance user representations for collaborative recommendations by capturing auxiliary information from neighbors in social networks.
Due to the superior representation learning capability in graphs, Graph Neural Networks (GNNs), such as LightGCN~\cite{he2020lightgcn}, GTN~\cite{fan2022graph}, and LTGNN~\cite{zhang2024ltgnn}, are proposed to capture high-order collaborative knowledge on user-item interaction graphs for enhancing the performance of recommender systems.
More specifically, GNNs take advantage of the graph-structured nature of user-item interactions and model similar user behavior patterns toward items through information propagation on user-item interaction graphs. 
For example, GraphRec~\cite{fan2019graph,fan2020graph} introduces a graph attention network-based framework to encode user-item interactions and user-user social relations for social recommendations. 
LightGCN~\cite{he2020lightgcn} is introduced to largely simplify the GNN-based recommendation methods by removing feature transformation and nonlinear activation, achieving state-of-the-art prediction performance for recommendations. 
In addition, GTN~\cite{fan2020graph} and LTGNN~\cite{zhang2024ltgnn} provide an improvement by capturing the adaptive reliability and high-order linear-time patterns of interactions, respectively. 

\subsection{LLM-based Recommender Systems}
With the rapid development of Large Language Models (LLMs), such as ChatGPT and GPT-4, notable milestones have been showcased for revolutionizing natural language processing techniques~\cite{ding2024survey,ding2024fashionregen,wei2022chain,fan2025computational}.
In particular, LLMs equipped with billion-scale parameters have exhibited unprecedented language understanding and generation ability, along with remarkable generalization capability and reasoning skills that facilitate LLMs to better generalize to unseen tasks and domains~\cite{chen2024exploring}.
Given the emerging trends and aforementioned advancements of LLMs, LLM-empowered recommender systems have drawn increasing attention from recent studies and demonstrated distinctive abilities for advancing recommender systems~\cite{zhao2024recommender,wang2024rethinking,wang2025knowledge,qu2025generative}. 
Notably, to harness the distinctive capabilities of LLMs for advancing recommender systems, existing studies have actively investigated various paradigms~\cite{zhao2023survey}, including pre-training, fine-tuning, and prompting, for adapting LLMs to recommendation tasks. 
For example, P5~\cite{geng2022recommendation} introduces an LLM-based recommendation model that unifies diverse recommendation tasks by multi-task prompt-based pre-training, which achieves impressive zero-shot generalization capability to unseen recommendation tasks with personalized prompts. 

Despite their effectiveness, most LLM-based recommendation methods still have an intrinsic limitation on indexing users and items' IDs in language models. 
The straightforward strategy, Independent Indexing (IID), involves directly allocating unique tokens to users and items.
However, it proves impractical and unfeasible for large-scale real-world recommendation systems, where the user and item populations can easily extend into the billions, significantly inflating the token vocabulary within LLMs.
As a natural solution, textual title indexing is proposed to utilize textual contents (e.g., titles and description) to tokenize items using LLMs' in-vocabulary tokens~\cite{gao2023chat,bao2023tallrec}, such as the example  ``\emph{Apple iPhone 15, 256 GB, black}". 
Moreover, P5~\cite{geng2022recommendation} and POD~\cite{li2023prompt} apply positional and whole-word embeddings to highlight the tokens representing items and users.
More recently, TIGER~\cite{rajput2023recommender} effectively condenses copious textual data into a few semantic IDs using a residual vector quantization method. By utilizing these quantized semantic IDs as tokens, TIGER is constructed as a Sequence-to-Sequence Transformer-based model for sequential recommendation tasks.
Although the uses of title indexing, whole-word embeddings, and semantic IDs can mitigate the issue of vocabulary explosion, they fail to capture high-order collaborative knowledge effectively and lack generalizability for recommending to unseen users or items.
To incorporate such valuable knowledge, several studies, including CoLLM~\cite{zhang2023collm}, LlaRA~\cite{liao2023llara}, and E4SRec~\cite{li2023e4srec}, borrow the conception of soft prompt and utilize exogenous tokens with continuous embeddings to represent users and items in LLM-based recommendations. 
However, the discrete nature of tokens in language models presents a challenge in achieving tight alignment of LLMs in recommendations when utilizing continuous representations.
Although META ID~\cite{huang2024improving} has suggested integrating collaborative knowledge into discrete tokens through clustering items' and users' representations derived from skip-gram models, a robust tokenizer remains unspecific to ensure the effectiveness of the quantization (i.e., clustering) procedure.
Additionally, most LLM-based RecSys suffer from time-consuming inference due to auto-regressive decoding and beam search.
To address these under-explored issues, this paper proposes a novel framework for LLM-based recommendations, \ourname{}, which not only introduces a generalizable ID tokenization strategy to capture high-order collaborative knowledge but also proposes a generative retrieval paradigm to generate top-$K$ items efficiently.
Unlike existing ID indexing methods that also utilize vector quantization, such as TIGER, our tokenizer features a parallel K-way structure and incorporates a masking operation for robust ID tokenization. Moreover, it is designed to capture insightful collaborative knowledge enriched with user-item interaction information.
\section{Conclusion}
\label{sec:conclusion}

While existing large language model-based recommendation methods achieve promising prediction performance, they fail to capture high-order collaborative knowledge and suffer from inferior generalization capability for tokenizing users\&items.
Additionally, the time-consuming inference remains an emerging challenge in LLM-based recommender systems. 
To tackle these shortcomings, we propose a novel approach, named \ourname{}, which not only introduces a generalizable user\&item ID tokenization strategy to capture high-order collaborative knowledge but also presents a generative retrieval paradigm for the efficient generation of top-$K$ items.
Particularly, a Masked Vector-Quantized Tokenizer (MQ-Tokenizer) is developed to tokenize users and items in LLM-based recommendations by incorporating high-order collaborative knowledge. 
Through comprehensive experiments on four distinct datasets, we demonstrated that our model can achieve state-of-the-art recommendation performance while also exhibiting the capacity to generalize to unseen users.

\section*{Acknowledgments}
The research described in this paper has been partially supported by the National Natural Science Foundation of China (project no. 62102335), General Research Funds from the Hong Kong Research Grants Council (project no. PolyU 15207322, 15200023, 15206024, and 15224524), internal research funds from Hong Kong Polytechnic University (project no. P0042693, P0048625, and P0051361). This work was supported by computational resources provided by The Centre for Large AI Models (CLAIM) of The Hong Kong Polytechnic University.

\bibliographystyle{IEEEtran}
\bibliography{references}

\begin{thebibliography}{10}
\providecommand{\url}[1]{#1}
\csname url@samestyle\endcsname
\providecommand{\newblock}{\relax}
\providecommand{\bibinfo}[2]{#2}
\providecommand{\BIBentrySTDinterwordspacing}{\spaceskip=0pt\relax}
\providecommand{\BIBentryALTinterwordstretchfactor}{4}
\providecommand{\BIBentryALTinterwordspacing}{\spaceskip=\fontdimen2\font plus
\BIBentryALTinterwordstretchfactor\fontdimen3\font minus \fontdimen4\font\relax}
\providecommand{\BIBforeignlanguage}[2]{{%
\expandafter\ifx\csname l@#1\endcsname\relax
\typeout{** WARNING: IEEEtran.bst: No hyphenation pattern has been}%
\typeout{** loaded for the language `#1'. Using the pattern for}%
\typeout{** the default language instead.}%
\else
\language=\csname l@#1\endcsname
\fi
#2}}
\providecommand{\BIBdecl}{\relax}
\BIBdecl

\bibitem{ning2024cheatagent}
L.-b. Ning, S.~Wang, W.~Fan, Q.~Li, X.~Xu, H.~Chen, and F.~Huang, ``Cheatagent: Attacking llm-empowered recommender systems via llm agent,'' in \emph{Proceedings of the 30th ACM SIGKDD Conference on Knowledge Discovery and Data Mining}, 2024, pp. 2284--2295.

\bibitem{chen2023fairly}
X.~Chen, W.~Fan, J.~Chen, H.~Liu, Z.~Liu, Z.~Zhang, and Q.~Li, ``Fairly adaptive negative sampling for recommendations,'' in \emph{Proceedings of the ACM Web Conference 2023}, 2023, pp. 3723--3733.

\bibitem{zhao2024recommender}
Z.~Zhao, W.~Fan, J.~Li, Y.~Liu, X.~Mei, Y.~Wang, Z.~Wen, F.~Wang, X.~Zhao, J.~Tang \emph{et~al.}, ``Recommender systems in the era of large language models (llms),'' \emph{IEEE Transactions on Knowledge and Data Engineering}, 2024.

\bibitem{fan2019graph}
W.~Fan, Y.~Ma, Q.~Li, Y.~He, E.~Zhao, J.~Tang, and D.~Yin, ``Graph neural networks for social recommendation,'' in \emph{WWW}, 2019, pp. 417--426.

\bibitem{koren2009matrix}
Y.~Koren, R.~Bell, and C.~Volinsky, ``Matrix factorization techniques for recommender systems,'' \emph{Computer}, vol.~42, no.~8, pp. 30--37, 2009.

\bibitem{he2020lightgcn}
X.~He, K.~Deng, X.~Wang, Y.~Li, Y.~Zhang, and M.~Wang, ``Lightgcn: Simplifying and powering graph convolution network for recommendation,'' in \emph{Proceedings of the 43rd International ACM SIGIR conference on research and development in Information Retrieval}, 2020, pp. 639--648.

\bibitem{fan2022graph}
W.~Fan, X.~Liu, W.~Jin, X.~Zhao, J.~Tang, and Q.~Li, ``Graph trend filtering networks for recommendation,'' in \emph{Proceedings of the 45th International ACM SIGIR Conference on Research and Development in Information Retrieval}, 2022, pp. 112--121.

\bibitem{brown2020language}
T.~Brown, B.~Mann, N.~Ryder, M.~Subbiah, J.~D. Kaplan, P.~Dhariwal, A.~Neelakantan, P.~Shyam, G.~Sastry, A.~Askell \emph{et~al.}, ``Language models are few-shot learners,'' \emph{Advances in neural information processing systems}, vol.~33, pp. 1877--1901, 2020.

\bibitem{zhao2023survey}
W.~X. Zhao, K.~Zhou, J.~Li, T.~Tang, X.~Wang, Y.~Hou, Y.~Min, B.~Zhang, J.~Zhang, Z.~Dong \emph{et~al.}, ``A survey of large language models,'' \emph{arXiv preprint arXiv:2303.18223}, 2023.

\bibitem{ning2025survey}
L.~Ning, Z.~Liang, Z.~Jiang, H.~Qu, Y.~Ding, W.~Fan, X.-y. Wei, S.~Lin, H.~Liu, P.~S. Yu \emph{et~al.}, ``A survey of webagents: Towards next-generation ai agents for web automation with large foundation models,'' \emph{arXiv preprint arXiv:2503.23350}, 2025.

\bibitem{gao2023chat}
Y.~Gao, T.~Sheng, Y.~Xiang, Y.~Xiong, H.~Wang, and J.~Zhang, ``Chat-rec: Towards interactive and explainable llms-augmented recommender system,'' \emph{arXiv preprint arXiv:2303.14524}, 2023.

\bibitem{geng2022recommendation}
S.~Geng, S.~Liu, Z.~Fu, Y.~Ge, and Y.~Zhang, ``Recommendation as language processing (rlp): A unified pretrain, personalized prompt \& predict paradigm (p5),'' in \emph{Proceedings of the 16th ACM Conference on Recommender Systems}, 2022, pp. 299--315.

\bibitem{bao2023tallrec}
K.~Bao, J.~Zhang, Y.~Zhang, W.~Wang, F.~Feng, and X.~He, ``Tallrec: An effective and efficient tuning framework to align large language model with recommendation,'' in \emph{Proceedings of the 17th ACM Conference on Recommender Systems}, 2023.

\bibitem{takase2020all}
S.~Takase and S.~Kobayashi, ``All word embeddings from one embedding,'' \emph{Advances in Neural Information Processing Systems}, vol.~33, pp. 3775--3785, 2020.

\bibitem{liao2023llara}
J.~Liao, S.~Li, Z.~Yang, J.~Wu, Y.~Yuan, X.~Wang, and X.~He, ``Llara: Aligning large language models with sequential recommenders,'' \emph{arXiv preprint arXiv:2312.02445}, 2023.

\bibitem{zhang2023collm}
Y.~Zhang, F.~Feng, J.~Zhang, K.~Bao, Q.~Wang, and X.~He, ``Collm: Integrating collaborative embeddings into large language models for recommendation,'' \emph{arXiv preprint arXiv:2310.19488}, 2023.

\bibitem{esser2021taming}
P.~Esser, R.~Rombach, and B.~Ommer, ``Taming transformers for high-resolution image synthesis,'' in \emph{Proceedings of the IEEE/CVF conference on computer vision and pattern recognition}, 2021, pp. 12\,873--12\,883.

\bibitem{ramesh2021zero}
A.~Ramesh, M.~Pavlov, G.~Goh, S.~Gray, C.~Voss, A.~Radford, M.~Chen, and I.~Sutskever, ``Zero-shot text-to-image generation,'' in \emph{International Conference on Machine Learning}.\hskip 1em plus 0.5em minus 0.4em\relax PMLR, 2021, pp. 8821--8831.

\bibitem{van2017neural}
A.~van~den Oord, O.~Vinyals, and K.~Kavukcuoglu, ``Neural discrete representation learning,'' in \emph{Proceedings of the 31st International Conference on Neural Information Processing Systems}, 2017.

\bibitem{wang2019neural}
X.~Wang, X.~He, M.~Wang, F.~Feng, and T.-S. Chua, ``Neural graph collaborative filtering,'' in \emph{ACM SIGIR}, 2019.

\bibitem{wang2024can}
H.~Wang, S.~Feng, T.~He, Z.~Tan, X.~Han, and Y.~Tsvetkov, ``Can language models solve graph problems in natural language?'' \emph{Advances in Neural Information Processing Systems}, vol.~36, 2024.

\bibitem{berthet2020learning}
Q.~Berthet, M.~Blondel, O.~Teboul, M.~Cuturi, J.-P. Vert, and F.~Bach, ``Learning with differentiable pertubed optimizers,'' \emph{Advances in neural information processing systems}, vol.~33, pp. 9508--9519, 2020.

\bibitem{lorberbom2019direct}
G.~Lorberbom, A.~Gane, T.~Jaakkola, and T.~Hazan, ``Direct optimization through argmax for discrete variational auto-encoder,'' \emph{Advances in neural information processing systems}, vol.~32, 2019.

\bibitem{bengio2013estimating}
Y.~Bengio, N.~L{\'e}onard, and A.~Courville, ``Estimating or propagating gradients through stochastic neurons for conditional computation,'' \emph{arXiv preprint arXiv:1308.3432}, 2013.

\bibitem{touvron2023llama}
H.~Touvron, T.~Lavril, G.~Izacard, X.~Martinet, M.-A. Lachaux, T.~Lacroix, B.~Rozi{\`e}re, N.~Goyal, E.~Hambro, F.~Azhar \emph{et~al.}, ``Llama: Open and efficient foundation language models,'' \emph{arXiv preprint arXiv:2302.13971}, 2023.

\bibitem{li2023mage}
T.~Li, H.~Chang, S.~Mishra, H.~Zhang, D.~Katabi, and D.~Krishnan, ``Mage: Masked generative encoder to unify representation learning and image synthesis,'' in \emph{Proceedings of the IEEE/CVF Conference on Computer Vision and Pattern Recognition}, 2023, pp. 2142--2152.

\bibitem{zhu2024beyond}
L.~Zhu, F.~Wei, and Y.~Lu, ``Beyond text: Frozen large language models in visual signal comprehension,'' in \emph{Proceedings of the IEEE/CVF Conference on Computer Vision and Pattern Recognition}, 2024, pp. 27\,047--27\,057.

\bibitem{zheng2023adapting}
B.~Zheng, Y.~Hou, H.~Lu, Y.~Chen, W.~X. Zhao, and J.-R. Wen, ``Adapting large language models by integrating collaborative semantics for recommendation,'' \emph{arXiv preprint arXiv:2311.09049}, 2023.

\bibitem{wang2024rethinking}
H.~Wang, X.~Liu, W.~Fan, X.~Zhao, V.~Kini, D.~Yadav, F.~Wang, Z.~Wen, J.~Tang, and H.~Liu, ``Rethinking large language model architectures for sequential recommendations,'' \emph{arXiv preprint arXiv:2402.09543}, 2024.

\bibitem{yang2020mixed}
J.~Yang, X.~Yi, D.~Zhiyuan~Cheng, L.~Hong, Y.~Li, S.~Xiaoming~Wang, T.~Xu, and E.~H. Chi, ``Mixed negative sampling for learning two-tower neural networks in recommendations,'' in \emph{Companion proceedings of the web conference 2020}, 2020, pp. 441--447.

\bibitem{cakir2019deep}
F.~Cakir, K.~He, X.~Xia, B.~Kulis, and S.~Sclaroff, ``Deep metric learning to rank,'' in \emph{Proceedings of the IEEE/CVF conference on computer vision and pattern recognition}, 2019, pp. 1861--1870.

\bibitem{mcfee2010metric}
B.~McFee and G.~Lanckriet, ``Metric learning to rank,'' 2010.

\bibitem{liu2009learning}
T.-Y. Liu \emph{et~al.}, ``Learning to rank for information retrieval,'' \emph{Foundations and Trends{\textregistered} in Information Retrieval}, vol.~3, no.~3, pp. 225--331, 2009.

\bibitem{zhang2024ltgnn}
J.~Zhang, R.~Xue, W.~Fan, X.~Xu, Q.~Li, J.~Pei, and X.~Liu, ``Linear-time graph neural networks for scalable recommendations,'' in \emph{Proceedings of the ACM on Web Conference 2024}, 2024, pp. 3533--3544.

\bibitem{liu2024lost}
N.~F. Liu, K.~Lin, J.~Hewitt, A.~Paranjape, M.~Bevilacqua, F.~Petroni, and P.~Liang, ``Lost in the middle: How language models use long contexts,'' \emph{Transactions of the Association for Computational Linguistics}, vol.~12, pp. 157--173, 2024.

\bibitem{kaneko2023reducing}
M.~Kaneko and N.~Okazaki, ``Reducing sequence length by predicting edit spans with large language models,'' in \emph{Proceedings of the 2023 Conference on Empirical Methods in Natural Language Processing}, 2023, pp. 10\,017--10\,029.

\bibitem{raffel2020t5}
C.~Raffel, N.~Shazeer, A.~Roberts, K.~Lee, S.~Narang, M.~Matena, Y.~Zhou, W.~Li, and P.~J. Liu, ``Exploring the limits of transfer learning with a unified text-to-text transformer,'' \emph{Journal of machine learning research}, vol.~21, no. 140, pp. 1--67, 2020.

\bibitem{Rendle2009BPRBP}
S.~Rendle, C.~Freudenthaler, Z.~Gantner, and L.~Schmidt-Thieme, ``Bpr: Bayesian personalized ranking from implicit feedback,'' in \emph{UAI}, 2009.

\bibitem{he2017neural}
X.~He, L.~Liao, H.~Zhang, L.~Nie, X.~Hu, and T.-S. Chua, ``Neural collaborative filtering,'' in \emph{Proceedings of the 26th international conference on world wide web}, 2017, pp. 173--182.

\bibitem{kang2018sasrec}
W.-C. Kang and J.~McAuley, ``Self-attentive sequential recommendation,'' in \emph{ICDM}.\hskip 1em plus 0.5em minus 0.4em\relax IEEE, 2018.

\bibitem{sun2019bert4rec}
F.~Sun, J.~Liu, J.~Wu, C.~Pei, X.~Lin, W.~Ou, and P.~Jiang, ``Bert4rec: Sequential recommendation with bidirectional encoder representations from transformer,'' in \emph{ACM CIKM}, 2019.

\bibitem{zhou2020s3}
K.~Zhou, H.~Wang, W.~X. Zhao, Y.~Zhu, S.~Wang, F.~Zhang, Z.~Wang, and J.-R. Wen, ``S3-rec: Self-supervised learning for sequential recommendation with mutual information maximization,'' in \emph{Proceedings of the 29th ACM international conference on information \& knowledge management}, 2020, pp. 1893--1902.

\bibitem{liu2021contrastive}
Z.~Liu, Y.~Chen, J.~Li, P.~S. Yu, J.~McAuley, and C.~Xiong, ``Contrastive self-supervised sequential recommendation with robust augmentation,'' \emph{arXiv preprint arXiv:2108.06479}, 2021.

\bibitem{hua2023index}
W.~Hua, S.~Xu, Y.~Ge, and Y.~Zhang, ``How to index item ids for recommendation foundation models,'' in \emph{Proceedings of the Annual International ACM SIGIR Conference on Research and Development in Information Retrieval in the Asia Pacific Region}, 2023, pp. 195--204.

\bibitem{li2023prompt}
L.~Li, Y.~Zhang, and L.~Chen, ``Prompt distillation for efficient llm-based recommendation,'' in \emph{ACM CIKM}, 2023.

\bibitem{rajput2023recommender}
S.~Rajput, N.~Mehta, A.~Singh, R.~Hulikal~Keshavan, T.~Vu, L.~Heldt, L.~Hong, Y.~Tay, V.~Tran, J.~Samost \emph{et~al.}, ``Recommender systems with generative retrieval,'' \emph{Advances in Neural Information Processing Systems}, vol.~36, pp. 10\,299--10\,315, 2023.

\bibitem{aloshchilov2019adamw}
I.~Loshchilov and F.~Hutter, ``Decoupled weight decay regularization,'' in \emph{Proceeding of the 7th International Conference on Learning Representations}, 2019.

\bibitem{lee2022autoregressive}
D.~Lee, C.~Kim, S.~Kim, M.~Cho, and W.-S. Han, ``Autoregressive image generation using residual quantization,'' in \emph{CVPR}, 2022.

\bibitem{fan2021attacking}
W.~Fan, T.~Derr, X.~Zhao, Y.~Ma, H.~Liu, J.~Wang, J.~Tang, and Q.~Li, ``Attacking black-box recommendations via copying cross-domain user profiles,'' in \emph{2021 IEEE 37th International Conference on Data Engineering (ICDE)}.\hskip 1em plus 0.5em minus 0.4em\relax IEEE, 2021, pp. 1583--1594.

\bibitem{qu2024ssd4rec}
H.~Qu, Y.~Zhang, L.~Ning, W.~Fan, and Q.~Li, ``Ssd4rec: a structured state space duality model for efficient sequential recommendation,'' \emph{arXiv preprint arXiv:2409.01192}, 2024.

\bibitem{fan2019deep_daso}
W.~Fan, T.~Derr, Y.~Ma, J.~Wang, J.~Tang, and Q.~Li, ``Deep adversarial social recommendation,'' in \emph{28th International Joint Conference on Artificial Intelligence (IJCAI-19)}.\hskip 1em plus 0.5em minus 0.4em\relax International Joint Conferences on Artificial Intelligence, 2019, pp. 1351--1357.

\bibitem{fan2018deep}
W.~Fan, Q.~Li, and M.~Cheng, ``Deep modeling of social relations for recommendation,'' in \emph{Proceedings of the AAAI Conference on Artificial Intelligence}, vol.~32, no.~1, 2018.

\bibitem{fan2019deep_dscf}
W.~Fan, Y.~Ma, D.~Yin, J.~Wang, J.~Tang, and Q.~Li, ``Deep social collaborative filtering,'' in \emph{Proceedings of the 13th ACM conference on recommender systems}, 2019, pp. 305--313.

\bibitem{fan2020graph}
W.~Fan, Y.~Ma, Q.~Li, J.~Wang, G.~Cai, J.~Tang, and D.~Yin, ``A graph neural network framework for social recommendations,'' \emph{IEEE Transactions on Knowledge and Data Engineering}, vol.~34, no.~5, pp. 2033--2047, 2020.

\bibitem{ding2024survey}
W.~Fan, Y.~Ding, L.~Ning, S.~Wang, H.~Li, D.~Yin, T.-S. Chua, and Q.~Li, ``A survey on rag meeting llms: Towards retrieval-augmented large language models,'' in \emph{Proceedings of the 30th ACM SIGKDD Conference on Knowledge Discovery and Data Mining}, 2024, pp. 6491--6501.

\bibitem{ding2024fashionregen}
Y.~Ding, Y.~Ma, W.~Fan, Y.~Yao, T.-S. Chua, and Q.~Li, ``Fashionregen: Llm-empowered fashion report generation,'' \emph{arXiv preprint arXiv:2403.06660}, 2024.

\bibitem{wei2022chain}
J.~Wei, X.~Wang, D.~Schuurmans, M.~Bosma, F.~Xia, E.~Chi, Q.~V. Le, D.~Zhou \emph{et~al.}, ``Chain-of-thought prompting elicits reasoning in large language models,'' \emph{Advances in neural information processing systems}, vol.~35, pp. 24\,824--24\,837, 2022.

\bibitem{fan2025computational}
W.~Fan, Y.~Zhou, S.~Wang, Y.~Yan, H.~Liu, Q.~Zhao, L.~Song, and Q.~Li, ``Computational protein science in the era of large language models (llms),'' \emph{arXiv preprint arXiv:2501.10282}, 2025.

\bibitem{chen2024exploring}
Z.~Chen, H.~Mao, H.~Li, W.~Jin, H.~Wen, X.~Wei, S.~Wang, D.~Yin, W.~Fan, H.~Liu \emph{et~al.}, ``Exploring the potential of large language models (llms) in learning on graphs,'' \emph{ACM SIGKDD Explorations Newsletter}, vol.~25, no.~2, pp. 42--61, 2024.

\bibitem{wang2025knowledge}
S.~Wang, W.~Fan, Y.~Feng, X.~Ma, S.~Wang, and D.~Yin, ``Knowledge graph retrieval-augmented generation for llm-based recommendation,'' \emph{arXiv preprint arXiv:2501.02226}, 2025.

\bibitem{qu2025generative}
H.~Qu, W.~Fan, and S.~Lin, ``Generative recommendation with continuous-token diffusion,'' \emph{arXiv preprint arXiv:2504.12007}, 2025.

\bibitem{li2023e4srec}
X.~Li, C.~Chen, X.~Zhao, Y.~Zhang, and C.~Xing, ``E4srec: An elegant effective efficient extensible solution of large language models for sequential recommendation,'' \emph{arXiv preprint arXiv:2312.02443}, 2023.

\bibitem{huang2024improving}
T.-J. Huang, J.-Q. Yang, C.~Shen, K.-Q. Liu, D.-C. Zhan, and H.-J. Ye, ``Improving llms for recommendation with out-of-vocabulary tokens,'' \emph{arXiv preprint arXiv:2406.08477}, 2024.

\end{thebibliography}

\vskip -0.35in

{\begin{IEEEbiography}[{\includegraphics[width=1in,height=1.25in,clip,keepaspectratio]{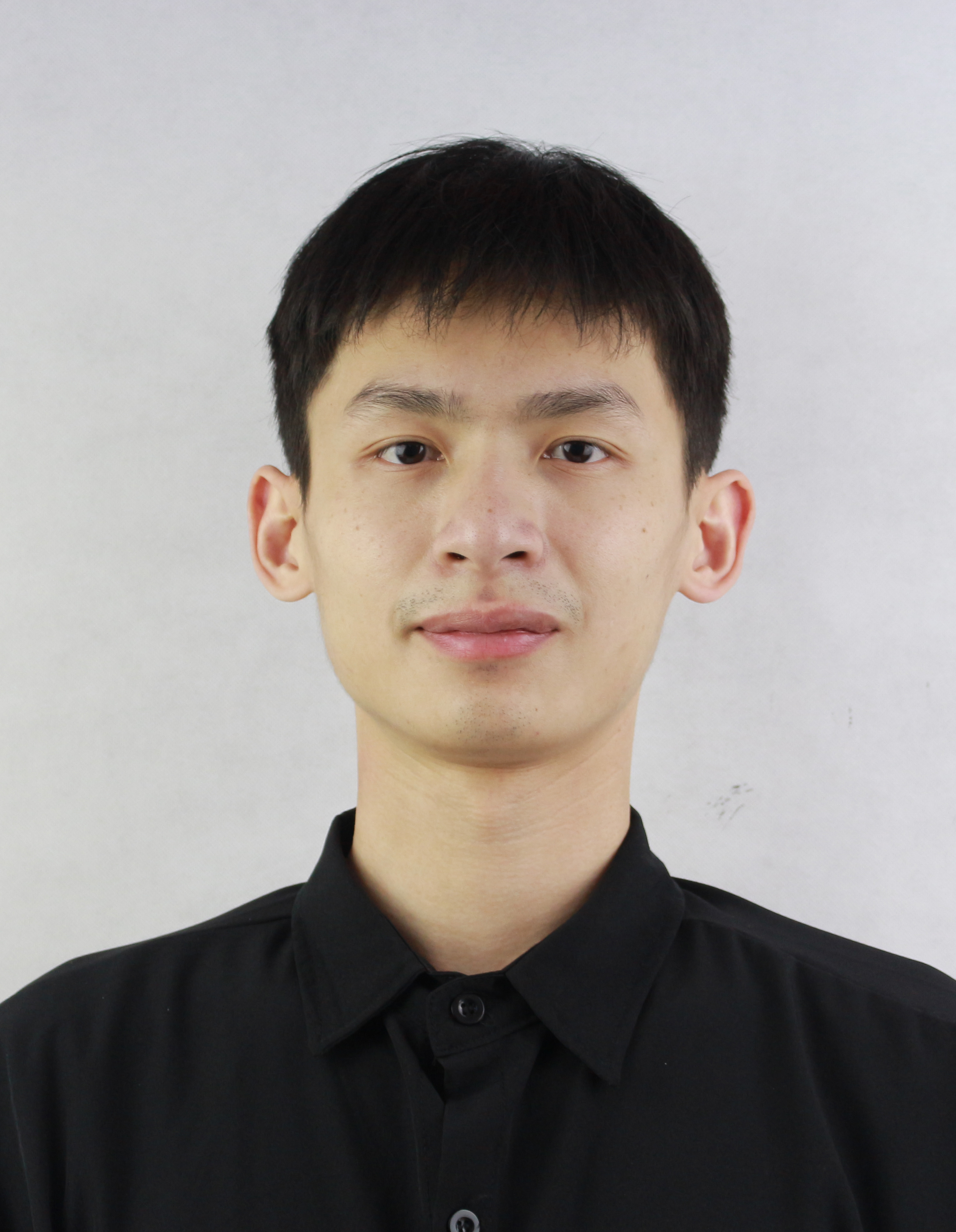}}]
{Haohao Qu} is currently a PhD student of the Department of Computing (COMP), The Hong Kong Polytechnic University (PolyU), under the supervision of Dr. Wenqi Fan and Prof. Qing Li. Before joining the PolyU, he received both a Master’s degree and a Bachelor’s degree from Sun Yat-sen University in 2022 and 2019, respectively. His research interest covers Recommender Systems and Large Language Models. For more information, please visit https://quhaoh233.github.io/page.
\end{IEEEbiography}

\vskip -0.35in

{\begin{IEEEbiography}[{\includegraphics[width=1in,height=1.25in,clip,keepaspectratio]{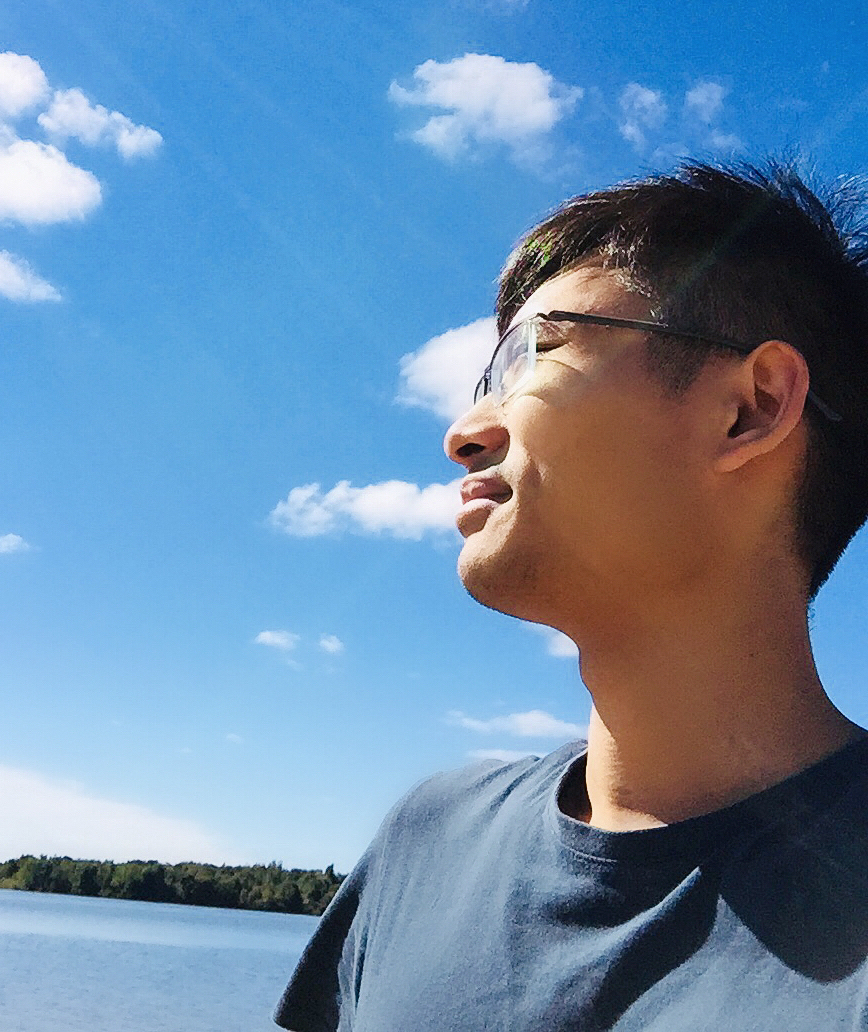}}]{Wenqi Fan} is an assistant professor of the Department of Computing (COMP) and the Department of Management and Marketing (MM) at The Hong Kong Polytechnic University (PolyU). He received his Ph.D. degree from the City University of Hong Kong (CityU) in 2020.
From 2018 to 2020, he was a visiting research scholar at Michigan State University (MSU). 
His research interests are in the broad areas of machine learning and data mining, with a particular focus on Recommender Systems, Graph Neural Networks, and Trustworthy Recommendations. He has published innovative papers in top-tier journals and conferences such as  TKDE, TIST, KDD, WWW, ICDE, NeurIPS, ICLR, SIGIR, IJCAI, AAAI, RecSys, WSDM, etc. 
He serves as top-tier conference (Area/Senior) Program Committee members and session chairs (e.g., ICML, ICLR, NeurIPS, KDD, WWW, AAAI, IJCAI, WSDM, EMNLP, ACL,  etc.), and journal reviewers (e.g., TKDE, TIST, TKDD, TOIS, TAI, etc.). 
More information about him can be found at https://wenqifan03.github.io.
\end{IEEEbiography}

\vskip -0.35in

{\begin{IEEEbiography}[{\includegraphics[width=1in,height=1.25in,clip,keepaspectratio]{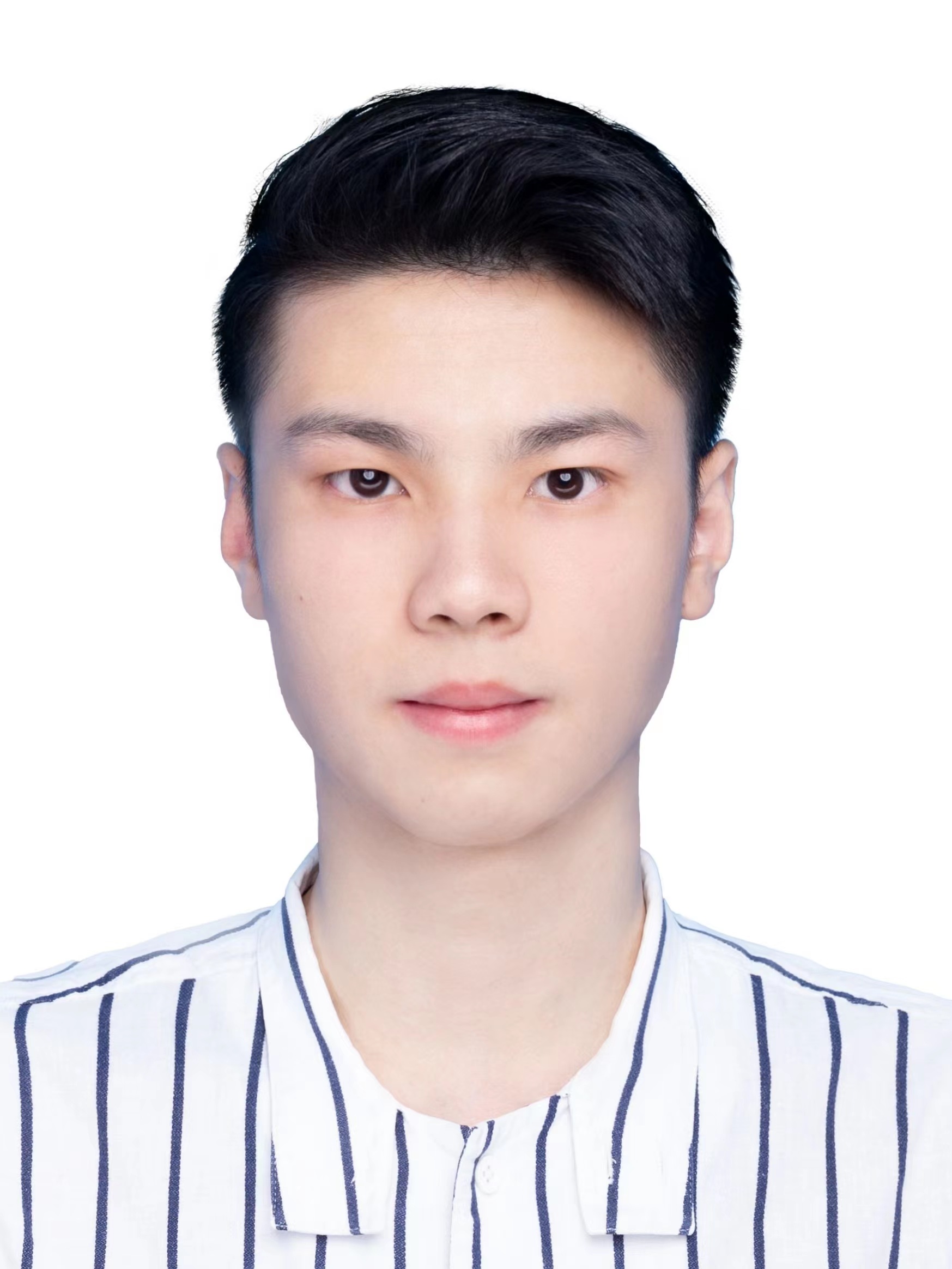}}]
{Zihuai Zhao} is currently a PhD student of the Department of Computing (COMP), Hong Kong Polytechnic University (PolyU), under the supervision of Prof. Qing Li and Dr. Wenqi Fan. Before joining the PolyU, he received both a Master’s degree (MPhil in Electrical Engineering) and a Bachelor’s degree (B.Eng. (Hons) in Electrical Engineering) from the University of Sydney in 2023 and 2020, respectively. His research interest covers Recommender Systems, Natural Language Processing, and Deep Reinforcement Learning. For more information, please visit https://scofizz.github.io/.
\end{IEEEbiography}

\vskip -0.35in
\begin{IEEEbiography}
[{\includegraphics[width=1in,height=1.25in,clip,keepaspectratio]{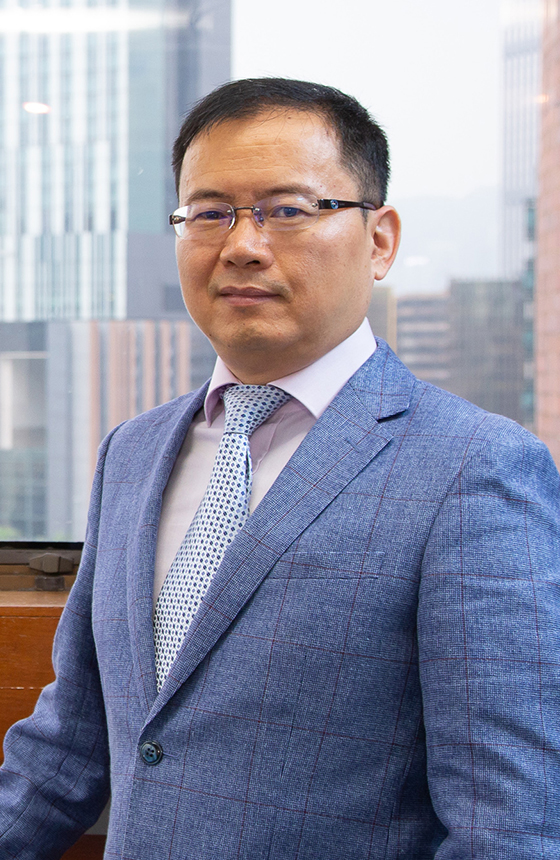}}]{Qing Li}
received the B.Eng. degree from Hunan University, Changsha, China, and the M.Sc. and Ph.D. degrees from the University of Southern California, Los Angeles, all in computer science.
He is currently a Chair Professor (Data Science) and the Head of the Department of Computing, the Hong Kong Polytechnic University. He is a Fellow of IEEE and IET, a member of ACM SIGMOD and IEEE Technical Committee on Data Engineering. 
His research interests include object modeling, multimedia databases, social media, and recommender systems. 
He has been actively involved in the research community by serving as an associate editor and reviewer for technical journals, and as an organizer/co-organizer of numerous international conferences. He is the chairperson of the Hong Kong Web Society, and also served/is serving as an executive committee (EXCO) member of IEEE-Hong Kong Computer Chapter and ACM Hong Kong Chapter. In addition, he serves as a councilor of the Database Society of Chinese Computer Federation (CCF), a member of the Big Data Expert Committee of CCF, and is a Steering Committee member of DASFAA, ER, ICWL, UMEDIA, and WISE Society. 
\end{IEEEbiography}
} 

\vfill
\end{document}